\documentclass[showpacs,pra,superscriptaddress,twocolumn]{revtex4-1}
\usepackage{pgf}
\usepackage[utf8]{inputenc}
\usepackage{amsmath}
\usepackage{braket}
\usepackage{amssymb}
\usepackage{amsmath}
\usepackage{enumitem} 
\usepackage{appendix}
\usepackage{graphicx}
\usepackage{ulem}
\usepackage{epstopdf}
\usepackage[pdftex,bookmarks,colorlinks]{hyperref}
\usepackage{hyperref}
\usepackage{amsmath}
\usepackage{amssymb}
\usepackage{braket}
\usepackage{ulem}
\newcommand{\Tr}[1]{\operatorname{Tr} #1}
\newcommand{\mean}[1]{\left\langle #1\right\rangle}
\newcommand{\nep}{\textrm{e}}

\definecolor{caribbeangreen}{rgb}{0.0, 0.8, 0.6}

\newcommand{\ud}{\mathrm{d}}
\newcommand\mf[1]{\textcolor{black}{#1}}

\newcommand{\kbar}{\mathchar'26\mkern-9mu k}
\newcommand{\opbdag}[1]{{\hat{b}^{\dagger}}_{#1}}
\newcommand{\opb}[1]{{\hat{b}^{\phantom \dagger}}_{#1}}

\begin{document}
\title{Chaos and subdiffusion in the infinite-range coupled quantum kicked rotors}
\author{Angelo Russomanno}
\affiliation{Max-Planck-Institut f\"ur Physik Komplexer Systeme, N\"othnitzer Stra{\ss}e 38, D-01187, Dresden, Germany}
\author{Michele Fava}
\affiliation{Rudolf Peierls Centre for Theoretical Physics, Clarendon Laboratory, University of Oxford, Oxford OX1 3PU, United Kingdom}
\author{Rosario Fazio}
\affiliation{Abdus Salam ICTP, Strada Costiera 11, I-34151 Trieste, Italy}
\affiliation{Dipartimento di Fisica, Universit{\`a} di Napoli ``Federico II'', Monte S. Angelo, I-80126 Napoli, Italy}
\normalem
\begin{abstract}
 We map the infinite-range coupled quantum kicked rotors over an infinite-range coupled interacting bosonic model. In this way we can apply exact diagonalization up to quite large system sizes and confirm that the system tends to ergodicity in the large-size limit. In the thermodynamic limit the system is described by a set of coupled Gross-Pitaevskii equations equivalent to an effective nonlinear single-rotor Hamiltonian. These equations give rise to a power-law increase in time of the energy with exponent $\gamma\sim 2/3$ in a \mf{wide} range of parameters. We explain this finding by means of a master-equation approach based on the noisy behaviour of the effective nonlinear single-rotor Hamiltonian \mf{and} on the Anderson localization of the single-rotor Floquet states. Furthermore, we study chaos by means of the largest Lyapunov exponent and find that it decreases towards zero for portions of the phase space with increasing momentum. Finally, we show that some stroboscopic Floquet integrals of motion of the noninteracting dynamics deviate from their initial values over a time scale related to the interaction strength according to the Nekhoroshev theorem. 
\end{abstract}
\maketitle
\section{Introduction}
%

The kicked rotor is a paradigmatic model in classical Hamiltonian dynamics. This simple model has been widely used to numerically study the development of chaos when integrability is broken, in accordance with the Kolmogorov-Arnold-Moser theorem~\cite{Chirikov1979263,lichtenberg1983regular,Berry_regirr78:proceeding}. In this single-degree-of-freedom model there can be weak chaos when the integrability breaking is small and there are still many manifolds of integrable-like dynamics (conserved tori) in phase space. In this case of weak chaos, the energy does not increase in time, while in the case of strong chaos the dynamics is ergodic and diffusive in phase space and the energy linearly increases in time. Considering the case of many coupled kicked rotors, on the opposite, there are not enough conserved tori and the dynamics eventually shows a diffusive behaviour with energy linearly increasing in time~\cite{Chirikov1979263,lichtenberg1983regular}. Nevertheless, the time scale after which this behaviour appears is very long and exponentially large with the inverse of the perturbation from integrability~\cite{PhysRevA.40.6130,konishi}. This is a particular case of a general theorem due to Nekhoroshev~\cite{Nekhoroshev1971,pettiland,notola}. So, a classical Hamiltonian system slightly perturbed from integrability shows a thermal ergodic behaviour after a long non-thermal transient, in strict analogy with the well-known quantum prethermalization~\cite{ue_da20}.

The single kicked rotor is very interesting also from the quantum point of view. Indeed this model shows a lack of correspondence between classical and quantum behaviour. Differently from other cases~\cite{Berry1_1977,PhysRevLett.51.943,Sred_PRE94,PhysRevA.34.591,Prosen_AJ,PhysRevLett.52.1,PhysRevLett.75.2300}, the kicked rotor shows a diffusive dynamics with linearly increasing energy in the strong classical regime which has no counterpart in the quantum domain. When the model is quantized due to quantum interference the classical energy increase stops after a transient.  This behaviour is \mf{known as} quantum dynamical localization~\cite{CHIRIKOV198877,Boris:rotor,chirikov}, \mf{and} has been interpreted as a dynamical disorder-free analog of Anderson localization~\cite{PhysRevLett.49.509,PhysRevA.29.1639,Delande_notes} and has also been experimentally observed~\cite{PhysRevLett.73.2974,PhysRevLett.75.4598}.

It is interesting to understand if dynamical localization persists in the case of many interacting rotors. \mf{It has been argued for different types of interactions, that for any set of parameters there is a threshold in system size beyond which the system becomes ergodic and the energy increases in time without a bound~\cite{Qin2017,kicked-rotors}}. 
So, in contrast with other models~\cite{QRKR,rylands,Michele_arxiv}, there is no dynamical localization in the many-body limit. Nevertheless, the system does not become equivalent to the classical model. 

This is clear in the case of infinite-range coupling where the thermodynamic-limit dynamics can be exactly computed~\cite{kicked-rotors} and one finds that the energy increases as a power law with exponent $\gamma<1$ (signaling subdiffusion in momentum space): Quantum effects make the classical linear energy increase slower. A power-law increase of the energy has also been observed in the quantum few-rotor interacting case~\cite{Notarnicola_2020} and in many non-linear generalizations of the single quantum kicked rotor~\cite{refId0,d4freq,PhysRevLett.100.094101} and related disordered models~\cite{PhysRevLett.70.1787,Nguyen2011,PhysRevLett.112.170603}. On the opposite, in the classical chaotic many-rotor case, the energy increases linearly in time~\cite{PhysRevA.40.6130,konishi}. One can observe a linear increase of the energy also in a single quantum kicked rotor if the kicking amplitude is modulated by a noise~\cite{klappauf1998observation}.

In this work we focus again on the quantum infinite-range coupled quantum kicked rotors. We restrict to the subspace even under all the site-permutation transformations,  we apply a method used before in another infinite-range coupled context~\cite{federica}, and map the model over a bosonic infinite-range interacting model over a lattice. From one side this fact allows us to apply exact diagonalization for larger system sizes and larger truncations with respect to what previously done. In this way we can probe ergodicity by means of the average level spacing ratio for larger system sizes $N$ and find further confirmation for the generalized tendency towards ergodicity for increasing $N$ previously demonstrated~\cite{kicked-rotors}.

This bosonic mapping is convenient also because allows to show that, in the limit $N\to \infty$, the model is described by a system of Gross-Pitaevskii equations. In the limit of vanishing interactions, these equations are equivalent to the Schr\"odinger equation of the single kicked rotor represented in the momentum basis. They are exact in the limit $N\to\infty$, and we can show that in general they are equivalent to the Schr\"odinger dynamics of the non-linear single kicked rotor effective Hamiltonian found in~\cite{kicked-rotors}. The energy increases in time as a power law with exponent $\gamma<1$ (sub-linear way) and the power-law exponent appears to be constant in a \mf{wide} range of parameters of the model and  consistent with the value $\gamma\sim 2/3$ (in agreement with~\cite{kicked-rotors}).

We can analytically explain the value of this exponent by considering the nonlinear modulation of the kick in the effective Hamiltonian as a noise. This allows us to write a master equation for the density matrix. We move to the Floquet basis~\cite{cohen1992atom} and apply a coarse graining in time as in~\cite{Russomanno_PRB11}; exploiting momentum-space Anderson localization of the Floquet states we get a diffusion equation for the momentum-eigenstates occupations. From that, we get a self-consistent differential equation for the energy expectation providing the power-law increase with exponent $2/3$. Moreover, we predict that the coarse-grained squared non-linear modulation of the kick depends on time as $t^{-1/3}$ and we numerically verify this fact. Our approach can be applied also to the case of a kicking modulated by a noise with properties invariant under time translations. This gives rise to a diffusive behaviour of the energy, in agreement with~\cite{klappauf1998observation}.

Our master-equation approach is possible thanks to the chaotic behaviour of the Gross-Pitaevskii equations.
In order to probe the chaoticity properties of these equations, we evaluate the largest Lyapunov exponent, which gives a measure of the rate of exponential divergence of nearby trajectories~\cite{Ott:book}. We see that, when we consider parts of the phase space with larger and larger momentum, the largest Lyapunov exponent decreases as a power law towards 0. So, for increasing momentum, the system is still chaotic but becomes asymptotically regular in the limit of infinite momentum. This is in agreement with  the results of the average level spacing ratio at finite size suggesting full ergodicity in the large-size limit. 

The largest Lyapunov exponent results are relevant also for finite $N\gg 1$. Here, due to Heisenberg principle, the relevant dynamical variables have an initial uncertainty of order $1/\sqrt{N}$. For finite $N$ we can use exponential divergence of nearby trajectories to show that the Gross-Pitaevskii equations are valid for a time increasing as $\log N$, with a coefficient equal to the inverse of the relevant Lyapunov exponent. So, in the limit of infinite momentum, the Gross-Pitaevskii equations tend to be valid for an infinite time, as one naively would expect, being in that limit the integrability-breaking interaction term vanishingly small compared with the undriven term. Moreover, the validity of the Gross-Pitaevskii equations for a time logarithmic in $N$ in the case of a chaotic dynamics is a fact generally valid in bosonic mean field dynamics. This is relevant for instance in the context of time crystals~\cite{Matus_2019}.

Chaos in this model \mf{obeys} the Nekhoroshev theorem~\cite{Nekhoroshev1971}. In the noninteracting case the dynamics is the Schr\"odinger dynamics of the single rotor which behaves in an integrable way and the resulting constants of motion of the stroboscopic dynamics are related to the Floquet states. Turning on the interaction, the Schr\"odinger dynamics becomes the nonlinear Gross-Pitaevskii one, integrability is broken and the Floquet constants of motion are no more conserved. They deviate in time from their initial value over a time scale exponential in the inverse of the integrability-breaking interaction term. 

The paper is organized as follows. In Section~\ref{model:sec} we introduce the model Hamiltonian and we perform the mapping to the infinite-range bosonic model. We describe all the details of the construction of this bosonic representation in Appendix~\ref{boson:sec}. In Sec.~\ref{ratio:sec} we perform exact diagonalization on this Hamiltonian, with an appropriate truncation, and by means of the average level spacing ratio we show the existence of a generalized tendency to ergodicity for increasing system size. In Sec.~\ref{limit:sec} we show that in the limit $N\to\infty$ the dynamics of the bosonic Hamiltonian is described by a system of Gross-Pitaevskii equations completely equivalent to a non-linear single-rotor effective Hamiltonian. In Sec.~\ref{subdiff:sec} we use these equations to numerically study the evolution of the energy. We find that it increases in time with a power law. We analytically explain the power-law exponent $\gamma=2/3$ valid in a \mf{wide} range of parameters by using a master-equation approach. In Sec.~\ref{lyapunov:sec} we study the Lyapunov exponent and show that the rate of exponential divergence of the trajectories tends to zero for increasing considered values of the momentum. In Sec.~\ref{nekhoroshev:sec} we show that the time scale over which the vanishing-perturbation integrals of motion deviate from their initial value obeys the Nekhoroshev theorem. 
\section{Hamiltonian and mapping to the bosonic model} \label{model:sec}
So, this is the quantum infinite-range coupled kicked rotor
\begin{equation}\label{H_a:eqn}
\hat{H}(t) = \frac{1}{2}\sum_{l=1}^N \hat{p}_l^2+V(\hat{\boldsymbol\theta})\delta_1(t)\,;
\end{equation}
where we define~\cite{chiri_vov} $\delta_1(t)\equiv\sum_{n=-\infty}^{+\infty}\delta (t-n)$ and
\begin{equation} \label{H_sr:eqn}
 V(\hat{\boldsymbol\theta}) \equiv 
   \kbar K\sum_{l=1}^N \cos \hat{\theta}_l 
 -\frac{\kbar K\epsilon}{2(N-1)} \sum_{l,\, l'} \cos (\hat{\theta}_l-\hat{\theta}_{l'})\,
\end{equation}
(here we have replaced the physical coupling $\mathcal{K}$ with $\kbar K$, in order to somewhat simplify the subsequent formulae). The commutation relations
\begin{equation}\label{com:eqn}
	[\hat\theta_l, \hat p_{l'}]=i\kbar\delta_{l,\,l'}
\end{equation}
are valid, with $\kbar$ related to the physical parameters of the Hamiltonian (one arrives at Eq.~\eqref{H_a:eqn} after an appropriate rescaling~\cite{kicked-rotors}). In all the paper we will focus on the stroboscopic dynamics, looking at the system in the instant $n^+$, that's to say immediately after the $n^{\rm th}$ kick (in the text we will omit the superscript $^+$). Most importantly this model is invariant under all the site-permutation transformations. The subspace even under all these transformations is therefore an eigenspace of the Hamiltonian, a point which will be crucial in the next section.

We restrict to the subspace even under all the permutation transformations. Using the methods explained in~\cite{federica} we get the effective Hamiltonian (see Appendix~\ref{boson:sec})
\begin{align}\label{hamo:eqn}
  &\hat{H}_b(t) = \frac{\kbar^2}{2}\sum_{m=-\infty}^\infty m^2\hat{n}_m+\frac{\kbar K}{2}\Big[\sum_{m=-\infty}^\infty \left(\opbdag{m}\opb{m+1}+{\rm H.~c.}\right)\nonumber\\
  &-\frac{\epsilon}{2(N-1)}\hspace{-0.1cm}\sum_{m,m'=-\infty}^\infty\hspace{-0.4cm}(\opbdag{m+1}\opb{m}\opbdag{m'}\opb{m'+1}+{\rm H.~c.})\Big]\delta_1(t)\,,
\end{align}
where $\opb{m}$ are bosonic operators obeying the commutation relations $[\opb{m}, \opbdag{m'}]=\delta_{m,\,m'}$, $[\opb{m}, \opb{m'}]=0$ and we define $\hat{n}_m\equiv \opbdag{m}\opb{m}$. The bosons obey the constraint $\sum_{m}\hat{n}_m=N$. Here $m$ mark the discrete single-particle momentum eigenvalues; with this mapping we represent the dynamics in terms of occupations of these levels. 
In the next subsection we are going to discuss the ergodicity properties of the model by means of the level statistics. 
%
\subsection{Average level spacing ratio} \label{ratio:sec}
This bosonic representation is quite convenient from a technical point of view because it is possible to apply exact diagonalization to the Hamiltonian Eq.~\eqref{hamo:eqn} for system sizes and truncations of the Hilbert space significantly larger than those considered in~\cite{kicked-rotors,simtesi}. A very important object in a periodically-driven dynamics is the time-evolution operator over one period which for the Hamiltonian in Eq.~\eqref{H_a:eqn} is
$$\hat{U}\equiv \nep^{-iV(\hat{\boldsymbol\theta})/\kbar}\nep^{-i\sum_{l=1}^N \hat{p}_l^2/(2\kbar)}\,.$$
We report the expression of $V(\hat{\boldsymbol\theta})$ and $\sum_{l=1}^N \hat{p}_l^2$ in the bosonic representation in Eq.~\eqref{mappings:eqn}. We further restrict to the subspace even under the $m\to -m$ symmetry~\cite{notesymm}. In this subspace, we compute $\hat{U}$, diagonalize it, and get the many-body Floquet quasienergies $\mu_\alpha$ as the phases of the eigenvalues of $\hat{U}$~\cite{Shirley_PR65,Samba}. Of course, this is possible only imposing a truncation to the Hilbert space, restricting to the states for which $|m|\leq M$. In order to probe ergodicity of the system, we can evaluate the average level spacing ratio $r$~\cite{Pal_PhysRevB10} defined as
\begin{equation}
  r = \frac{1}{\mathcal{N}_M-2}\sum_{\alpha=1}^{\mathcal{N}_M-2}\frac{\min(\mu_{\alpha+2}-\mu_{\alpha+1},\mu_{\alpha+1}-\mu_\alpha)}{\max(\mu_{\alpha+2}-\mu_{\alpha+1},\mu_{\alpha+1}-\mu_\alpha)}
\end{equation}
where the quasienergies are restricted to the first quasienergy Brillouin zone~\cite{Russomanno_PRL12,rudner} $[-\pi,\pi]$ (they are periodic of period $2\pi$) and taken in increasing order. $\mathcal{N}_M$ is the dimension of the truncated Hilbert space. It is important that on the subspace even under all the permutations we have imposed the further constraint of being even under the $m\to -m$ symmetry. In this way we are restricting to an irreducible invariant subspace of the Hamiltonian, a condition required in order that the level spacing distribution (and the related ratio $r$) is a meaningful ergodicity indicator~\cite{Berry_LH84}. When the driven system is ergodic, {\em i.e.} locally thermalizing with $T=\infty$~\cite{Lazarides_PRE14}, the Floquet operator $\hat{U}$ belongs to the circular orthogonal ensemble (COE) of symmetric unitary matrices (because of the time-reversal invariance)~\cite{Rigol_PRX14,Haake:book,eynard2018random,Notarnicola_2020}. In this case, the level spacing distribution is of the COE type and the average level spacing ratio acquires the value $r_{\rm COE}\simeq 0.5269$. A level spacing distribution of the Poisson type corresponds to an integrable dynamics~\cite{Berry_PRS76} and is observed for instance in the single kicked rotor which is dynamically localized~\cite{Boris:rotor,kicked-rotors} and behaves in an integrable-like way breaking the classical ergodicity (one can contstruct infinite integrals of motion local in momentum space which deeply affect energy absorption). It corresponds to an average level spacing ratio $r_P\simeq 0.386$.

\begin{figure}
 \begin{center}
  \begin{tabular}{c}
    \includegraphics[width=8cm]{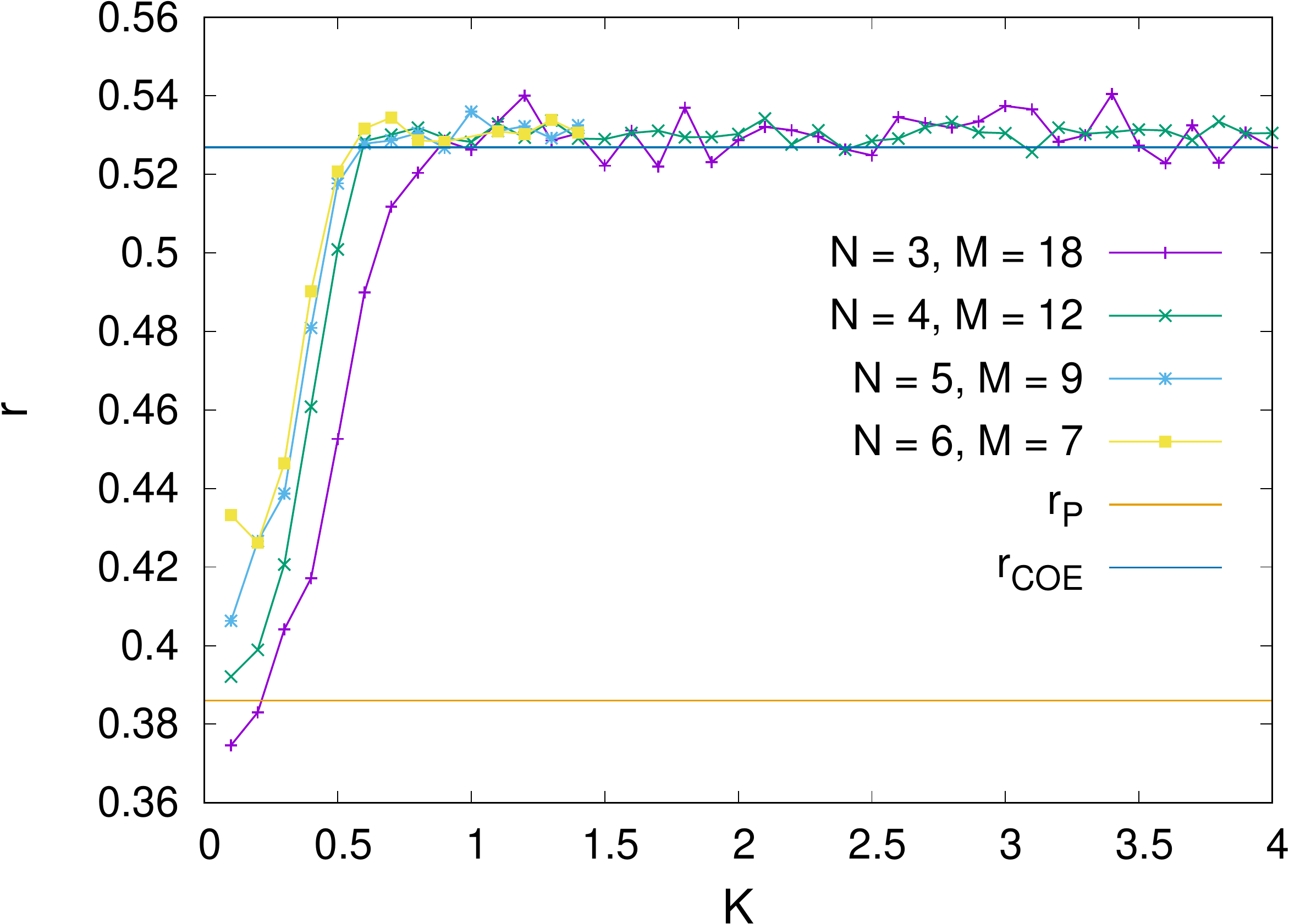}
  \end{tabular}
 \end{center}
 \caption{$r$ versus $K$ for different values of $N$. Numerical parameters: $\epsilon=0.52$, $\kbar = 1.7$.}
    \label{rivoli:fig}
\end{figure}

We show $r$ versus $K$ for a fixed $\epsilon$ in Fig.~\ref{rivoli:fig}. For each $N$ we choose $M$ so that $r$ has attained convergence (fixing $N$, we see a quite fast convergence of $r$ with the truncation $M$ of the Hilbert space). We see that from a certain $K$ on, $r$ attains the COE value: The system becomes ergodic. For increasing $N$ the values of $r$ generically increases and $r$ attains the COE value for smaller and smaller values of $K$. This suggests a tendency towards ergodicity for increasing size. This is in agreement with the analytical predictions of~\cite{kicked-rotors} of a completely ergodic system in the limit $N\to\infty$. 
\section{Gross-Pitaevskii equations in the $N\to\infty$ limit} \label{limit:sec}
\subsection{Derivation}\label{derivation:sec}
We start by considering the limit $N\to\infty$ where the dynamics is described by effective classical equations. Everything is based on the definition $\hat{\beta}_m\equiv\opb{m}/\sqrt{N}$ and on the observation that 
\begin{equation} \label{comm:eqn}
  [\hat{\beta}_m,\hat{\beta}_{m'}^\dagger]=\delta_{m,\,m'}/N\,,
\end{equation}
 so in the limit $N\to\infty$, the $\hat{\beta}_m$ behave as classical variables. We write the Heisenberg equations for the $\hat{\beta}_m$~\cite{noteb}
$$
  i\kbar\frac{\ud}{\ud t}\hat{\beta}_{m\,H}(t)=[\hat{\beta}_{m\,H}(t),\,\hat{H}_b(t)]
$$
and exploit the fact that in the limit $N\to\infty$ the $\hat{\beta}_m$ are uncorrelated, due to the vanishing commutators in this limit [Eq.~\eqref{comm:eqn}]. By defining $\beta_m(t)=\braket{\psi(t)|\hat{\beta}_m|\psi(t)}$, in the $N\to\infty$ limit we get from the Heisenberg equations the following system of Gross-Pitaevskii equations
\begin{align} \label{evolama:eqn}
  &i\frac{\ud}{\ud t}\beta_m(t)=\frac{\kbar}{2}m^2\beta_m(t) + \frac{K}{2}\delta_1(t)\bigg[\beta_{m+1}(t)+\beta_{m-1}(t)\nonumber\\
  &-\epsilon\Big(\sum_{m'}\beta_{m'}^*(t)\beta_{m'+1}(t)\Big)\beta_{m-1}(t)\nonumber\\
  &-\epsilon\Big(\sum_{m'}\beta_{m'}(t)\beta_{m'+1}^*(t)\Big)\beta_{m+1}(t)\bigg]\,.
\end{align}
The energy per rotor of the unkicked part of the model can be written as
\begin{equation}
  e(t) = \frac{\kbar^2}{2}\sum_{m}m^2|\beta_m(t)|^2\,.
\end{equation}
We see that for $\epsilon=0$ this system of equations is equivalent to the Schr\"odinger equation for the wave function of a single rotor in the momentum-basis representation~\cite{Haake:book, Stockmann:book}. 

The Gross-Pitaevskii equations can be obtained from the effective classical Hamiltonian
{\small
\begin{align}
  &\mathcal{H}(\{\beta_m,\beta_m^*\},t)=\sum_{m=-\infty}^\infty \frac{\kbar}{2}m^2|\beta_m(t)|^2+\frac{K}{2}\Big[\sum_{m=-\infty}^\infty (\beta_{m}^*\beta_{m+1}\nonumber\\
  &+{\rm H.~c.})-\frac{\epsilon}{2}\hspace{-0.2cm}\sum_{m,m'=-\infty}^\infty\hspace{-0.4cm}(\beta_{m+1}^*\beta_{m}\beta_{m'}^*\beta_{m'+1}+{\rm H.~c.})\Big]\delta_1(t)\,.
\end{align}
}
In order to get the Gross-Pitaevskii equations, one writes $\frac{\ud}{\ud t}\beta_m=-\{\beta_m,\,\mathcal{H}\}$ where $\{\cdots\}$ are the Poisson brackets~\cite{Arnold:book}, and the elementary Poisson brackets (needed to evaluate all the other ones) are
\begin{equation}\label{brackets:eqn}
  \{\beta_m,\,\beta_n^*\}=i\delta_{m\,n}\,\quad\text{and}\quad \{\beta_m,\,\beta_n\}=0\quad\forall\;n,\,m\,.
\end{equation}
These Poisson brackets will have a relevant role in the analysis related to the Nekhoroshev theorem in Sec.~\ref{nekhoroshev:sec}.
%
%

In the limit $M\to\infty$ we can actually map the semiclassical model Eq.~\eqref{evolama:eqn} into the self-consistent single-particle model studied in~\cite{kicked-rotors}.
In order to do that, we rewrite Eq.~\eqref{evolama:eqn} in a sort of continuum-limit approximation. Identifying $x\equiv m$ and $\psi(x)\equiv\beta_m$, expanding to all orders in $1/m$ as in~\cite{Sciolla_2}, we get
\begin{widetext}
\begin{align} \label{evolama1:eqn}
  &i\kbar\frac{\ud}{\ud t}\psi(x,t)=\frac{\kbar^2}{2}x^2\psi(x,t)
  + \frac{\kbar K}{2}\delta_1(t)\bigg[2\left(\cosh\left(\frac{\ud}{\ud x}\right)-1\right)\psi(x,t)\nonumber\\
  &-\epsilon\bigg(\int_{-\infty}^\infty\psi^*(x',t)\exp\left(\frac{\ud}{\ud x}\right)\psi(x',t)\ud x'\bigg)\exp\left(-\frac{\ud}{\ud x}\right)\psi(x,t)\nonumber\\
 &-\epsilon\bigg(\int_{-\infty}^\infty\psi^*(x',t)\exp\left(\frac{\ud}{\ud x}\right)\psi(x',t)\ud x'\bigg)^*\exp\left(\frac{\ud}{\ud x}\right)\psi(x,t)\bigg]\,,
\end{align}
\end{widetext}
where we have exploited that $\int_{-\infty}^\infty|\psi(x,t)|^2=1$ and used the formal relation $\psi(x+1)=\exp\left(\frac{\ud}{\ud x}\right)\psi(x)$, involving all the perturbative orders~\cite{Sciolla_2}. Introducing the appropriate coordinate and momentum operators
\begin{align}
  \hat{\theta}&=i\frac{\ud}{\ud x}\,\nonumber\\
  \hat{p}&=\kbar x\,,
\end{align}
such that $[\hat{\theta},\hat{p}]=i\kbar$,~\cite{notrasfo} we can rewrite this formula as
\begin{align} \label{evolama2:eqn}
  &i\kbar\frac{\ud}{\ud t}\ket{\psi_t}=\hat{H}_{\rm eff}(t)\ket{\psi_t}\,.
\end{align}
with the effective self-consistent Hamiltonian
\begin{align}\label{effe0:eqn}
  \hat{H}_{\rm eff}(t)&\equiv\frac{\hat{p}^2}{2} + \kbar K\delta_1(t)\bigg[\cos(\hat{\theta})-{\epsilon}\braket{\psi_t|\cos\hat{\theta}|\psi_t}\cos\hat{\theta}\nonumber\\
  &-{\epsilon}\braket{\psi_t|\sin\hat{\theta}|\psi_t}\sin\hat{\theta}\bigg]\,.
\end{align}
\mf{This is exactly the effective mean-field Hamiltonian following Eq.~\eqref{H_a:eqn}, found in~\cite{kicked-rotors} through a much more involuted analysis. We can easily see that, if the initial state is symmetric under $\hat{\theta}\mapsto-\hat{\theta}$, this symmetry will be preserved during the time-evolution, so that $\braket{\psi_t|\sin\hat{\theta}|\psi_t}=0$. Consequently
\begin{equation}\label{Ft:eqn}
  F(t)=\sum_m\beta_m^*(t)\beta_{m+1}(t)=\braket{\psi_t|\cos\hat{\theta}|\psi_t}-i\braket{\psi_t|\sin\hat{\theta}|\psi_t}\,
\end{equation}
is purely real. Thus,} the effective self-consistent Hamiltonian Eq.~\eqref{effe:eqn} acquires the form
\begin{align}\label{effe:eqn}
  \hat{H}_{\rm eff}(t)&\equiv\frac{\hat{p}^2}{2} + \kbar K\delta_1(t)\left[1-{\epsilon}F(t)\right]\cos\hat{\theta}\,.
\end{align}
This effective Hamiltonian will be very relevant for us in the next section, \mf{where} we will use it as the starting point to construct a master-equation model for explaining the energy subdiffusion with exponent $2/3$.
\section{Energy subdiffusion}\label{subdiff:sec}
\begin{figure}
 \begin{center}
  \begin{tabular}{c}
    \includegraphics[width=8cm]{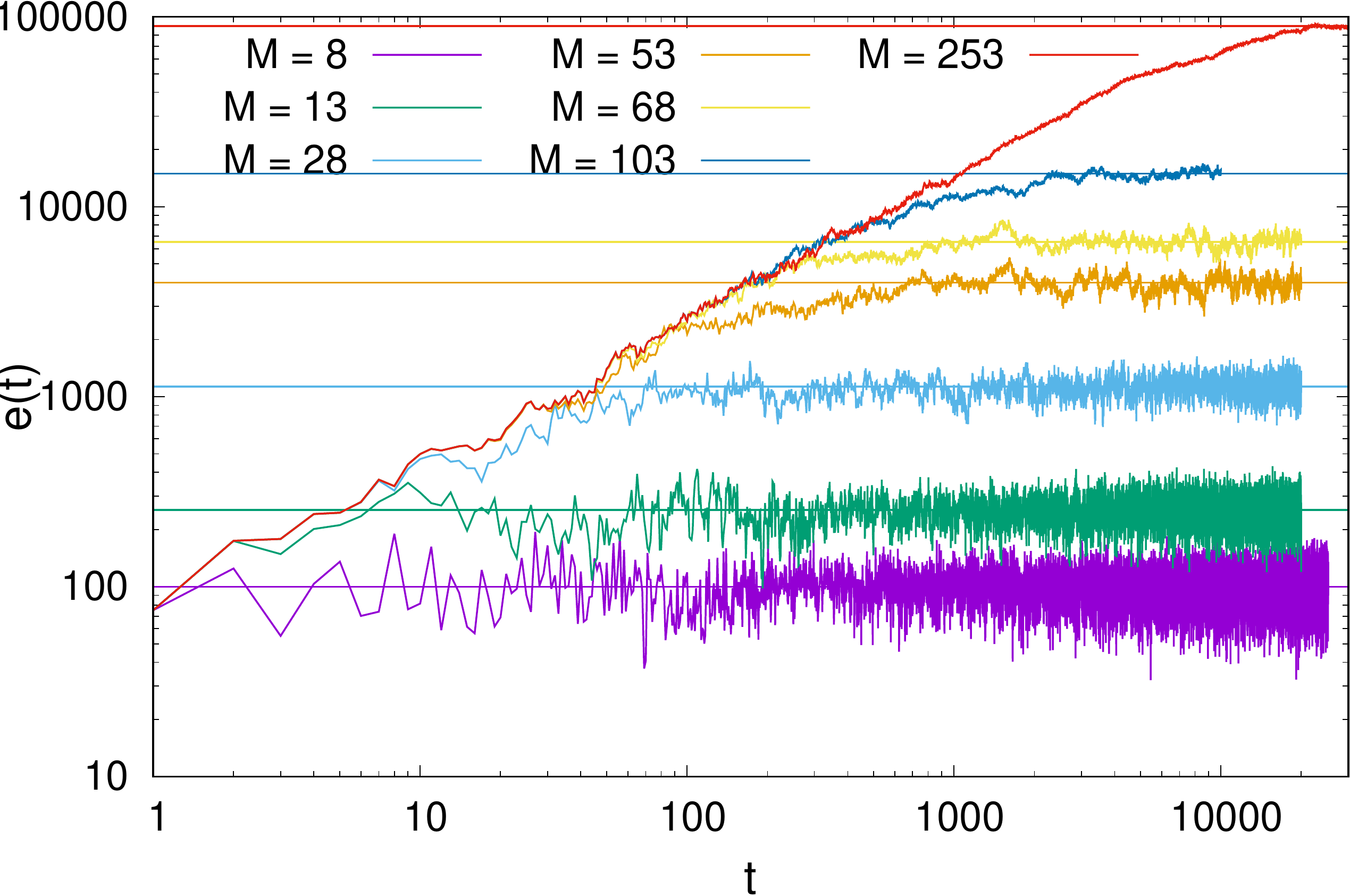}
  \end{tabular}
 \end{center}
 \caption{Stroboscopic evolution of $e(t)$ versus $t$ with truncation (bilogarithmic plot). We plot also the $T=\infty$ value in the truncated subspace (horizontal lines). We first see a power-law increase and then a convergence to the $T=\infty$ value. Without a truncation the power-law increase would last forever. Numerical parameters: $K=6$, $\epsilon=0.52$, $\kbar = 1.7$.}
    \label{semievo:fig}
\end{figure}

%
We numerically solve equations~\eqref{evolama:eqn} by means of a fourth order Runge-Kutta method with adaptive time step scheme~\cite{NumericalRecipesF,notevole}. We initialize them in the symmetrized $m=0$ state for the rotors, therefore we choose $\beta_m(0)=\left\{\begin{array}{ll}1&\text{for}\quad m=0\\0&\text{otherwise}\end{array}\right.$. and we can compare the results with those of the mean field analysis performed in a different way in~\cite{kicked-rotors}. In order to implement them we have to impose a truncation, fixing some $M>0$ and restricting to the values of $m$ such that $m\leq M$. We show some plots of $e(t)$ versus $t$ with stroboscopic time ($t$ is integer) for an interacting case with $\epsilon\neq 0$ in Fig.~\ref{semievo:fig}. The horizontal line corresponds to the $T=\infty$ value of the unkicked energy in the truncated subspace. This quantity can be readily evaluated as
\begin{align}
  e_{T=\infty}(M)&=\frac{\kbar^2}{2M+1}\sum_{m=1}^M m^2\nonumber\\
   &=\frac{\kbar^2}{2M+1}\left(\frac{M^3}{3}+\frac{M^2}{2}+\frac{M}{6}\right)\,.
\end{align}
In Fig.~\ref{semievo:fig} we can see convergence towards $e_{T=\infty}$, and therefore thermalization. 

\begin{figure}
 \begin{center}
  \begin{tabular}{c}
    \includegraphics[width=7.5cm]{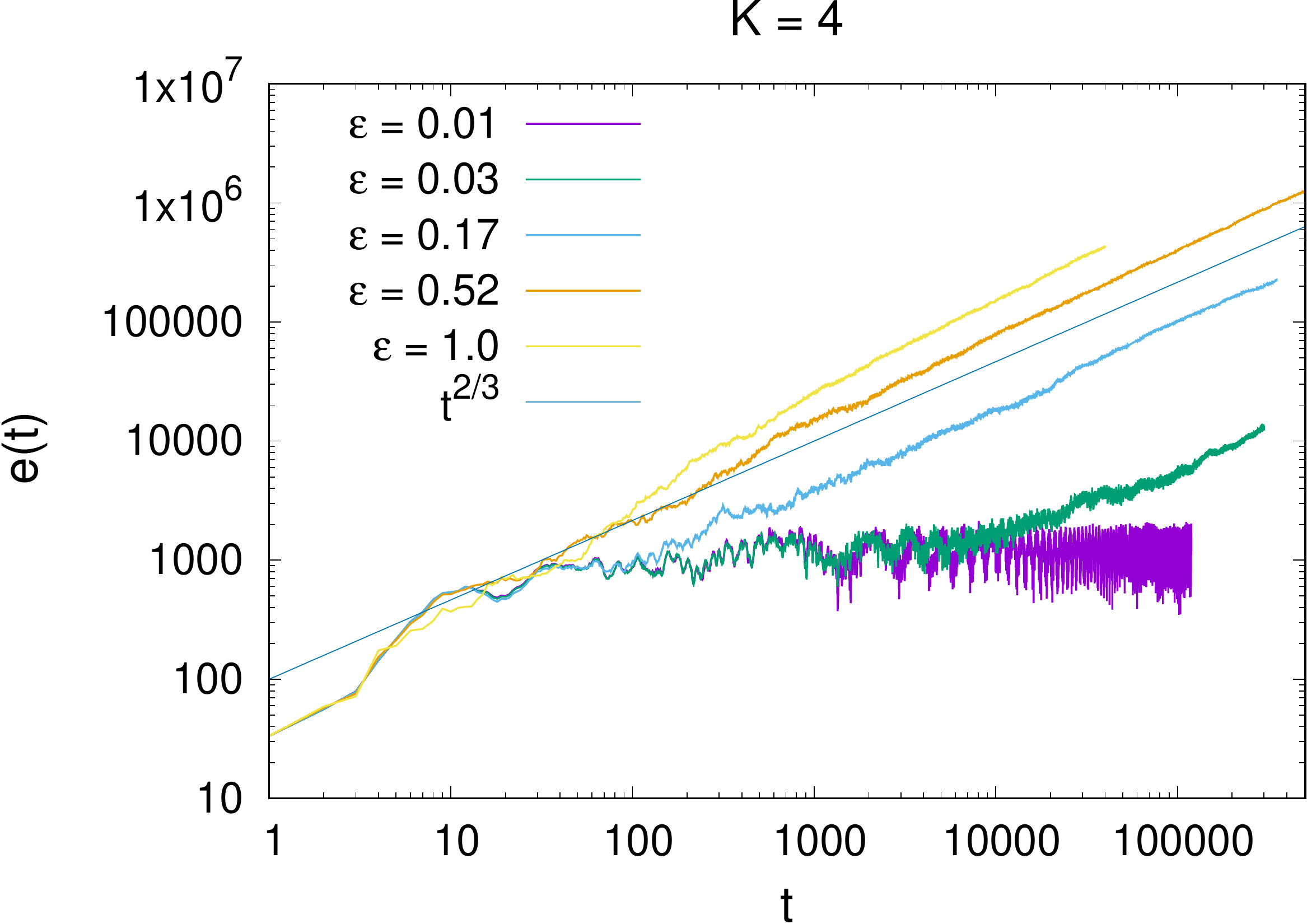}\\\vspace{0.cm}\\
    \includegraphics[width=7.5cm]{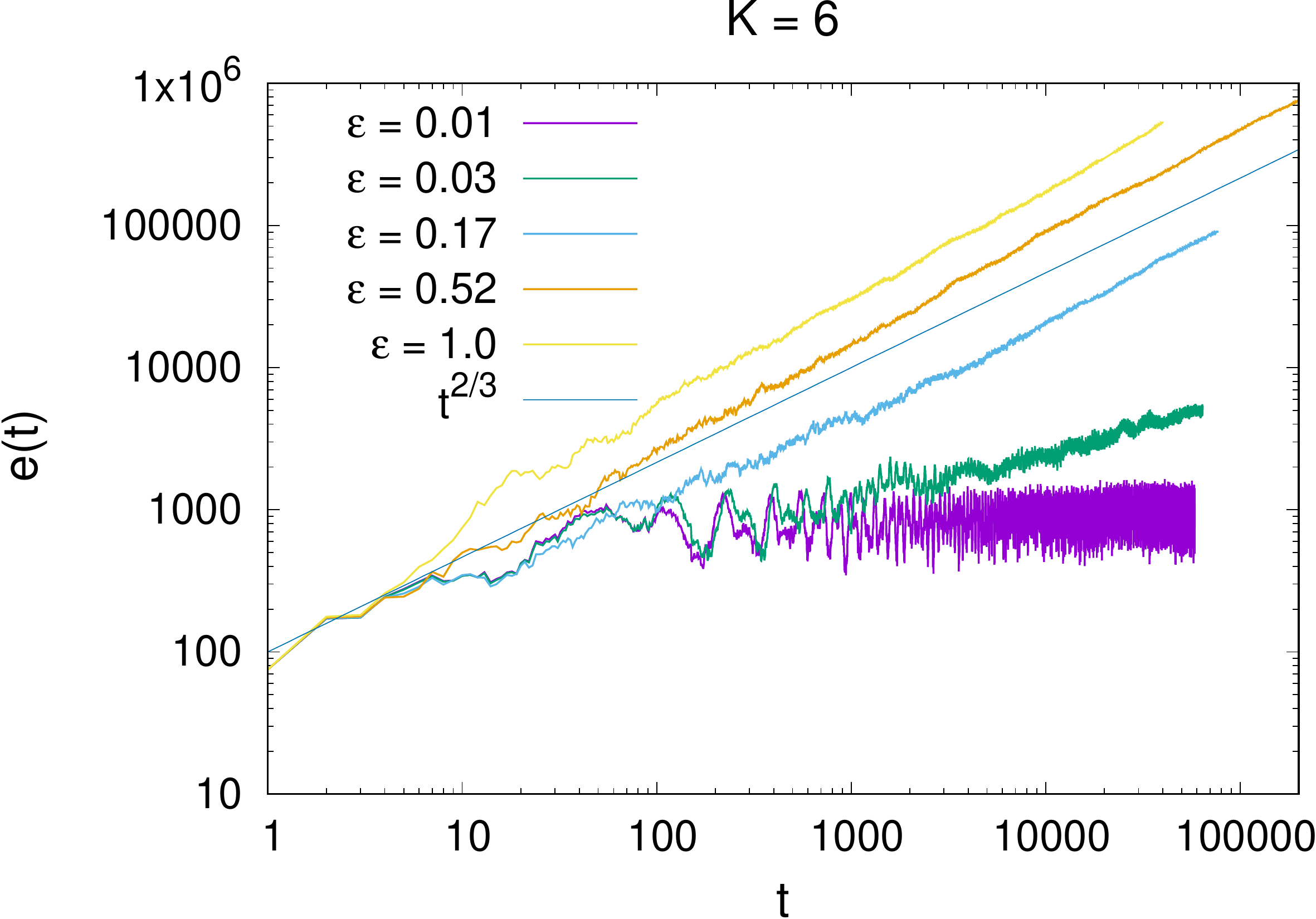}\\\vspace{0.cm}\\
    \includegraphics[width=7.5cm]{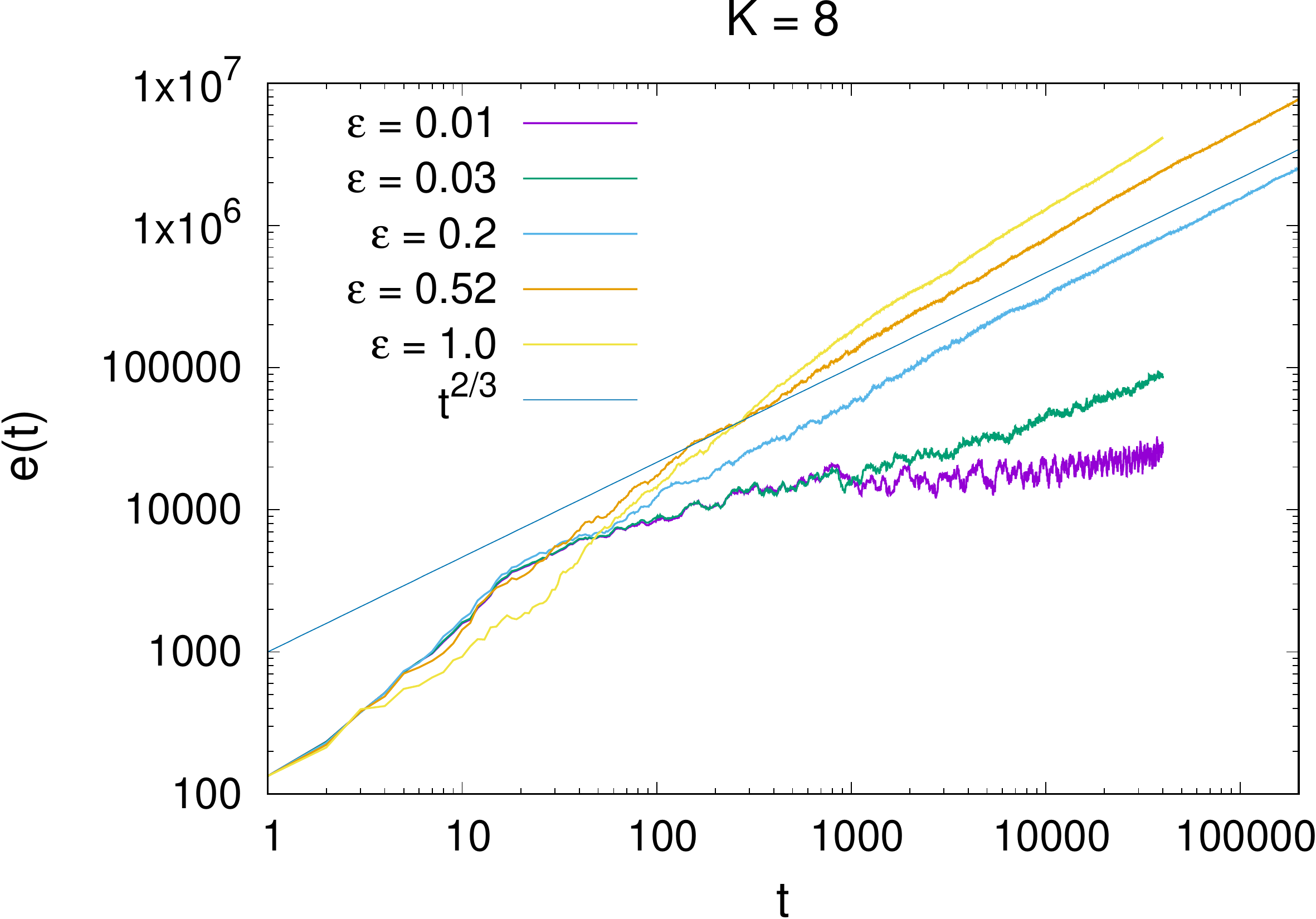}\\
  \end{tabular}
 \end{center}
 \caption{Stroboscopic evolution of $e(t)$ versus $t$ with truncation $M=20153$ for different values of the parameters (bilogarithmic plot). Notice the power-law increase which at long times tends towards a behaviour consistent with the law $e(t)\sim t^{2/3}$ when $\epsilon\in[0.1,1]$ (for $K=4$ this occurs also for $\epsilon=0.03$). Other numerical parameters: $\kbar = 1.7$.}
    \label{semievo1:fig}
\end{figure}

Most importantly, we see a power-law increase in time of the energy, still in agreement with the findings of~\cite{kicked-rotors}. We can see this power-law increase until saturation sets in, but with $M\to\infty$ it would last forever, as one can easily convince by looking at Fig.~\ref{semievo:fig}. We show some examples of power-law increase for different choices of parameters in Fig.~\ref{semievo1:fig}. The curves with $\epsilon\in[0.1,1]$ are consistent at long times with a power-law increase of the form 
$$
  e(t)\sim t^{\gamma}\quad {\rm with} \quad\gamma\sim 2/3\,.
$$
This numerical result was already found in~\cite{kicked-rotors} (see Figs.~9 and~10 of that paper). So, in a \mf{wide} range of $\epsilon$, the exponent of the power law appears to be independent of the choice of the parameters. In the following we provide an analytical argument for explaining this finding.
%

\mf{Before deriving the equation giving rise to momentum subdiffusion, we explain the rationale behind our analysis. The key observation is that energy absorption is purely driven by the fluctuation of the self-consistent field $F(t)$, in fact, if $F(t)$ were constant, the system would be in a dynamical-localized phase with an asymptotically constant energy. We can then expect that the evolution with the effective Hamiltonian Eq.~\eqref{effe0:eqn} gives a $\frac{d}{dt}\langle \hat{p}^2\rangle_t$ proportional to the variance of $F(t)$ over time --- as we will show in Sec.~\ref{master-eq:sec}. Furthermore, as the momentum-range 
$$m_t=\sqrt{\langle \hat{p}^2\rangle_t}/\kbar$$ grows in time, the variance of $F(t)$ decreases towards $0$ as $1/m_t$ (see Sec.~\ref{F-statistics:sec}). As a consequence the diffusion in momentum is slowed down as the wave-function spreads, giving rise to the subdiffusive behaviour observed above.
}

\subsection{\mf{Statistical properties of $F(t)$}}\label{F-statistics:sec}

\begin{figure}
 \begin{center}
  \begin{tabular}{c}
    \includegraphics[width=8cm]{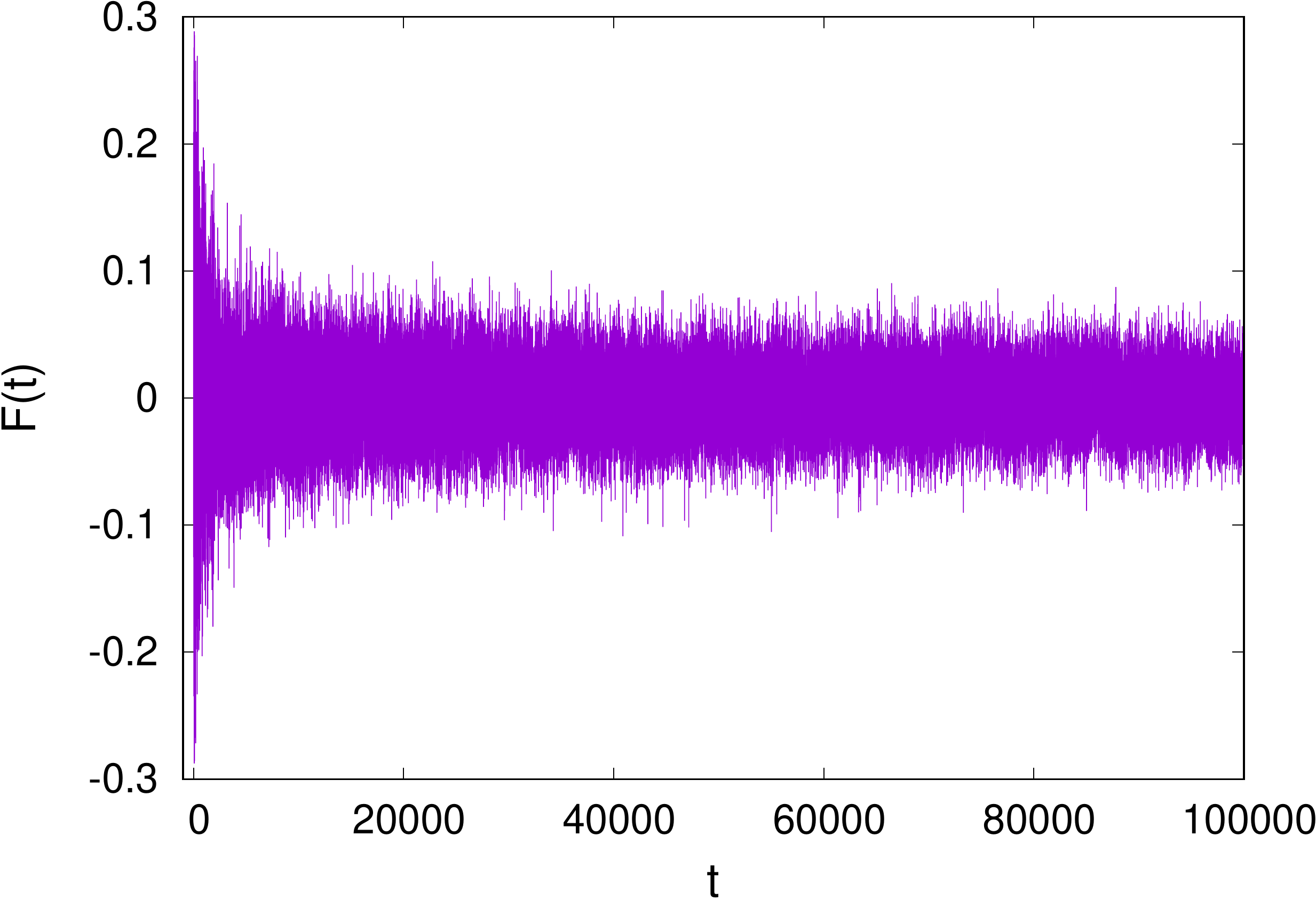}
  \end{tabular}
 \end{center}
 \caption{Example of evolution of $F(t)$. Notice the chaotic behaviour symmetric around 0. Numerical parameters: $K=4$, $\epsilon=0.52$, $\kbar = 1.7$, $M=20153$.}
    \label{semievon:fig}
\end{figure}

	While the non-linear equation Eq.~\eqref{evolama:eqn} is hard to characterize analytically, a simplifying assumption is to consider $F(t)$ as an effective noise. This is reasonable because, as a consequence of the chaotic diynamics in Eq.~\eqref{evolama:eqn}, $F(t)$ shows random oscillations symmetric around 0, as we can see in Fig.~\ref{semievon:fig}.
	Before starting our analysis we discuss which minimal set of self-consistency properties should the effective noise have in order to adequately describe $F(t)$.
	
	Firstly, looking at Eq.~\eqref{Ft:eqn}, we can split the sums over $m$ into pieces, each of one including $O(\xi_{\text{loc}})$ adjacent $m$s, where $\xi_{\text{loc}}$ is the single-rotor localization length (in momentum space). It is now reasonable to assume that the contribution of each of these pieces will be largely uncorrelated with the other ones. Then, when the state is spread over a momentum range $m_t\gg\xi_{\text{loc}}$, by the central limit theorem, the sum $F(t)$ will be distributed like a Gaussian with mean zero.
	
	To estimate the variance of the Gaussian noise, we use the normalization condition $\sum_m |\beta_m|^2=1$ to infer that the magnitude of $|\beta_m|$ scales like $1/\sqrt{m_t}$ for $m\lesssim m_t$, and is approximately zero otherwise. Then summing over a region of length $2\xi_{\text{loc}}$ around $\bar{m}$, such that $|\bar{m}|\lesssim m_t$, we approximately have
	\begin{equation}
		\text{Var}\left[\sum_{|m-\bar{m}|<\xi_{\text{loc}}} \beta_m^*(t)\beta_{m+1}(t) \right]\sim \left(\frac{\xi_{\text{loc}}}{m_t}\right)^2
	\end{equation}
	assuming perfect correlation inside that region. Considering, instead, pieces from different regions as uncorrelated, we have that
	\begin{equation}
		\text{Var}\left[F(t)\right] \sim \frac{m_t}{\xi_{\text{loc}}} \left(\frac{\xi_{\text{loc}}}{m_t}\right)^2 \sim \frac{\xi_{\text{loc}}}{m_t}
	\end{equation}
	Estimating the localization length in terms of the single-rotor parameter (see e.g. Ref.~\cite{Delande_notes}) we then have
	\begin{equation}\label{var-F-estimate:eq}
		\text{Var}\left[F(t)\right] \sim \frac{K^2}{ m_t}.
	\end{equation}
	Finally, since the evolution of $\boldsymbol{\beta}$ is chaotic (see Sec.~\ref{lyapunov:sec}), we will assume that the autocorrelation of $F(t)$ decays in time over a finite scale $\tau$.
	
	These assumptions are consistent with the numerical solution of the equation of motion. We show an example in Fig.~\ref{semievon:fig}, where we can see that the random oscillations are symmetric around 0 and $F(t)$ is short-range correlated in time.

\subsection{\mf{Effective master equation}}\label{master-eq:sec}

	We start by considering the self-consistent Hamiltonian Eq.~\eqref{effe:eqn} and we define
%
  $\hat{H}_0(t)\equiv \frac{\hat{p}^2}{2} + \kbar K\delta_1(t)\cos\hat{\theta}$, 
  $\hat{V} \equiv \cos\hat{\theta}$. 
%
First of all we move to the interaction representation defining $\hat{U}_0(t) \equiv \overleftarrow{\mathcal{T}}\exp\left( -(i/\kbar) \int^t dt' \hat{H}_0(t') \right)$, and $\hat{\mathcal{O}}_I(t)\equiv\hat{U}_0^\dagger(t)\hat{\mathcal{O}}\hat{U}_0(t)$, $\ket{\psi(t)}_I\equiv\hat{U}_0^\dagger(t)\ket{\psi(t)}$ for a generic operator $\hat{\mathcal{O}}$. (We have defined here the time-ordered-integral exponential as $\overleftarrow{\mathcal{T}}\exp$). In the interaction representation Eq.~\eqref{evolama2:eqn} takes the form
	\begin{equation}\label{inter:eqn}
		\frac{d}{dt} \ket{\psi(t)}_I =-i\epsilon K\delta_1(t) F(t) \hat{V}_I(t) \ket{\psi(t)}_I.
	\end{equation}
%
%
%
Treating $F(t)$ as a Gaussian noise, we can derive from Eq.~\eqref{inter:eqn} a master equation similar to the ones resulting from the coupling to an environment~\cite{cohen1992atom}. We write first the evolution equation for the density matrix $\hat{\rho}_I^\psi(t)\equiv\ket{\psi(t)}_I\!\bra{\psi(t)}$
\begin{equation}\label{interaction:eqn}
  i\frac{\ud }{\ud t}\hat{\rho}_I^\psi(t)=-\epsilon K \delta_1(t) F(t)[\hat{V}_I(t),\,\hat{\rho}_I^\psi(t)]\,.
\end{equation}
%
Integrating Eq.~\eqref{interaction:eqn} once we get
\begin{equation}
  \hat{\rho}_I^\psi(t) = \hat{\rho}_I^\psi(0) + i\epsilon K \int_0^t\delta_1(t')F(t')[\hat{V}_I(t'),\,\hat{\rho}_I^\psi(t')]\ud t'\,.
\end{equation}
Substituting this formula into Eq.~\eqref{interaction:eqn} and averaging over the Gaussian noise (the assumption of Gaussianity allows to neglect the correlations between $\hat{V}_I$ and $\hat{\rho}_I^\psi$) we get
\begin{widetext}
\begin{align}\label{prima:eqn}
 &\frac{\ud }{\ud t}\hat{\rho}_I(t)=-\epsilon^2 K^2\int_0^t\braket{F(t')F(t)}\delta_1(t)\delta_1(t')\left[\hat{V}_I(t),\,[\hat{V}_I(t'),\,\hat{\rho}_I(t')]\right]\ud t'\,,
\end{align}
\end{widetext}
where $\braket{\cdots}$ marks the average over the Gaussian ensemble and we have defined $\hat{\rho}_I(t)=\braket{\hat{\rho}_I^\psi(t)}$. 
%
%
%
\mf{At this point we apply a coarse-graining average over a mesoscopic time-scale $\Delta t\gg \tau,\,1$. Furthermore, we assume that averages of functions of $F(t)$ over time can be substituted with the corresponding averages over the Gaussian ensemble producing the noise (see Sec.~\ref{F-statistics:sec}). We expect this step to be valid as long as the system is effectively ergodic on the mesoscopic time-scale $\Delta t$, which is reasonable in this case, given that the system is chaotic (see Sec.~\ref{lyapunov:sec}).
Thus, we have}
\begin{equation}
  \braket{F(t')F(t)}\simeq\overline{F^2}(t)g(t-t')
\end{equation}
where $\overline{(\cdots)}(t)$ marks the average over the coarse-graining time interval $[t-\Delta t,\,t+\Delta t]$ and $g(t-t')$ is a function which decays to 0 when $|t-t'|>\tau$.
Integrating over $\ud t'$ and applying the coarse-graining average, the $\delta_1$ functions disappear and give rise to an average over integer times, multiple of the kicking period (stroboscopic times). The kicking period $1$ is much smaller than $\Delta t$, the resolution in time after coarse graining, then we can approximate the average over stroboscopic times with an average over continuous times. Moreover, assuming $\tau\lesssim\Delta t$, we can substitute $t'$ with $t$ because the function $g(t-t')$ acts as a delta function at the level of time resolution given by $\Delta t$. Defining 
\begin{equation}\label{sissi:eqn}
  \sigma(t)\equiv g\epsilon^2 K^2\, \overline{F^2}(t)\,,
\end{equation}
where the coefficient $g$ results from the integration of $g(t-t')$ and putting everything together, we get
\begin{equation}\label{evope1:eqn}
 \frac{\ud }{\ud t}\overline{\hat{\rho}_I}(t)=-\sigma(t)\overline{\left[\hat{V}_I(t),\,[\hat{V}_I(t),\,\hat{\rho}_I(t)]\right]}\,,
\end{equation}
%
where, due to the presence of the $\delta_1(t)$, the coarse-graining average results being over discrete values of time ($t_1$ is integer) and we have approximated the coarse-grained derivative as $\frac{ \hat{\rho}_I(t+\Delta t)-\hat{\rho}_I(t-\Delta t)}{2\Delta t}\simeq\frac{\ud}{\ud t}\overline{\hat{\rho}_I}(t)$ because we are interested in phenomena occurring over timescales larger than $\Delta t$. Moreover, $\sigma(t)$ is already coarse grained and the coarse-graining average does not affect it.
%
%
%

\mf{Finally, since we are interested in the asymptotic properties at large times, we can assume that $m_t\gg1$. Consequently, by Eq.~\eqref{var-F-estimate:eq}, we can assume that $\sigma(t)\ll 1$. This assumption is crucial for our analysis since it allows us to make a separation-of-timescales approximation~\cite{cohen1992atom}.} With this assumption, we can see from Eq.~\eqref{evope1:eqn} that the noise induces significant changes in $\hat{\rho}_I(t)$ over a time scale $t_r$ much larger than the typical order-1 timescale $t_0$ of the dynamics induced by $\hat{H}_0$. So, if we take $t_0\ll\Delta t\ll t_r$, $\hat{\rho}_I(t)$ in Eq.~\eqref{evope1:eqn} is not affected by the coarse-graining average. In essence, $\hat{\rho}_I(t)$ can be approximated by its coarse-grained average, $\hat{\rho}_I(t)\simeq \overline{\hat{\rho}_I}(t)$. This fact will have important consequences in the following analysis of subdiffusion.

%

\subsection{Subdiffusion}\label{subdiffusion:sec}
In order to derive subdiffusion, we need to consider the dynamics of the operator $\hat{O}=\ket{m}\bra{m}$ where $\ket{m}$ is an eigenstate of the operator $\hat{p}$ with eigenvalue $\kbar m$ and $m\in\mathbb{Z}$. Its expectation is the occupation probability of the state $\ket{m}$ and we will eventually find a diffusion equation for these occupations.

We can expand this operator in the basis of the noninteracting Floquet modes~\cite{Shirley_PR65,Samba}. They are defined in terms of the noninteracting Floquet states which are solutions of the noninteracting Schr\"odinger equation $i\kbar\partial_t\ket{\psi(t)}=\hat{H}_0(t)\ket{\psi(t)}$ which are periodic up to a phase $\ket{\psi_j(t)}=\nep^{-i\mu_j t}\ket{\phi_j(t)}$. The periodic part $\ket{\phi_j(t)}$ are the Floquet modes. We have defined $\hat{U}_0(t)$ as the time-evolution operator of $\hat{H}_0$ from time 0 to time $t$, and we find
\begin{equation}\label{ultron:eqn}
  \hat{U}_0(t)\ket{\phi_j(0)}=\nep^{-i\mu_j t}\ket{\phi_j(t)}\,.
\end{equation}
So, we get the expression
\begin{equation}\label{oexp:eqn}
  \hat{O}\equiv \sum_{i\,j}\mathcal{O}_{i\,j}(t)\ket{\phi_i(t)}\bra{\phi_j(t)}\,,
\end{equation}
where $\mathcal{O}_{i\,j}(t)\equiv\braket{\phi_j(t)|\hat{\mathcal{O}}|\phi_i(t)}$ is a periodic quantity. Now let us write its expectation at time $t$ in the interaction representation as
\begin{align}
  \mathcal{O}(t)&=\Tr[\hat{U}_0^\dagger(t)\,\hat{\mathcal{O}}\,\hat{U}_0\hat{\rho}_I(t)]\nonumber\\
   &=\sum_{i\,j}\mathcal{O}_{i\,j}(t)\nep^{i(\mu_i-\mu_j)t}\braket{\phi_j(0)|\hat{\rho}_I(t)|\phi_i(0)}
\end{align}
where we have used Eqs.~\eqref{ultron:eqn} and~\eqref{oexp:eqn}. We apply the coarse-graining average. Thanks to the separation of timescales, it does not act on $\hat{\rho}_I(t)$. It acts over many periods and, assuming that the $\mu_j$ are incommensurate with the driving frequency $2\pi$~\cite{Russomanno_PRB11}, we get
\begin{equation}
  \overline{\mathcal{O}}(t)=\sum_{i}\overline{\mathcal{O}_{i\,i}}\braket{\phi_i(0)|\hat{\rho}_I(t)|\phi_i(0)}\,.
\end{equation}
Being $\mathcal{O}_{i\,j}(t)$ periodic, $\overline{\mathcal{O}_{i\,i}}$ does not depend on the coarse-grained time. 
 Using Eq.~\eqref{evope1:eqn} and applying the cyclic property of trace, we get
\begin{align}\label{Om:eqn}
  &\frac{\ud}{\ud t}\overline{\mathcal{O}}(t)=\nonumber\\
  &-\sigma(t)\sum_i\overline{\mathcal{O}_{i\,i}}\Tr\Big\{\overline{\big[[\ket{\phi_i(0)}\bra{\phi_i(0)},\hat{V}_I(t)],\,\hat{V}_I(t)\big]}\hat{\rho}_I(t)\Big\}\,.
\end{align}
Expanding in the Floquet-mode basis $\hat{V}=\sum_{k\,l}V_{k\,l}(t)\ket{\phi_k(t)}\bra{\phi_l(t)}$ [as in Eq.~\eqref{oexp:eqn}] we obtain
\begin{equation}\label{Om1:eqn}
  \frac{\ud}{\ud t}\overline{\mathcal{O}}(t)=-2\sigma(t)\sum_{i\,k\neq i}\overline{\mathcal{O}_{i\,i}}\,\overline{|V_{k\,i}|^2}\braket{\phi_k(0)|\hat{\rho}_I(t)|\phi_k(0)}\,
\end{equation}
(we report the detailed derivation from Eq.~\eqref{Om:eqn} in Appendix~\ref{app:der}). Dynamical localization implies Anderson localization of the Floquet modes $\ket{\phi_k(0)}$ in the momentum basis $\{\ket{m}\}$~\cite{PhysRevLett.49.509,PhysRevA.29.1639,kicked-rotors}. Moreover, $\hat{V}$ is local in momentum space because it only connects $\ket{m}$ with $\ket{m+1}$. These two facts imply that $\overline{|V_{k\,i}|^2}$ is local (or short range) and is nonvanishing only if $|j-k|$ is smaller than some localization length $\lambda$. At this point we can substitute $\hat{O}=\ket{m}\bra{m}$ and define $p_m(t)\equiv\overline{\ket{m}\bra{m}}(t)=\braket{m|\hat{\rho}_I(t)|m}$. Concerning this quantity, Anderson localization of Floquet states has another important consequence. If we make a coarse-graining average in momentum space, then $\braket{\phi_k(0)|\hat{\rho}_I(t)|\phi_k(0)}$ can be approximated by $\braket{m'|\hat{\rho}_I(t)|m'}$ for some $m'$, and we get
	\begin{align}\label{equocan2:eqn}
     \frac{d}{dt} p_m(t)\simeq
  -\sigma(t) \sum_{m'}C_{m,\,m'}p_{m'}(t)\,,
	\end{align}
where $C_{m\,m'}$ is a coupling local in $m$ and $m'$.
%
%
In order to better specify the form of this local coupling, we consider that, thanks to the coarse graining in momentum space, $m$ can be assumed continuous. A constraint comes from the normalization condition $\sum_m p_m(t)=1$, which is automatically enforced by writing the right-hand side of Eq.~\eqref{equocan2:eqn} as a total derivative in $m$ 
	\begin{equation}\label{baroccolo:eqn}
		\frac{d}{dt} p_m(t) = -\sigma(t)\, \partial_m\, \mathcal{J}_m(\{p_{m'}(t)\})\,,
	\end{equation}
where $\mathcal{J}_m(\{p_{m'}(t)\})$ is linear in the $p_{m'}(t)$.
%
%
%
Due to the locality of $C_{m\,m'}$, we can apply to $\mathcal{J}_m(\{p_{m'}(t)\})$ a gradient expansion
	\begin{equation}
		\mathcal{J}_m(\{p_{m'}\}) = C(m) p_m- D(m)\, \partial_m p_m + \text{higher derivatives}.
	\end{equation}
	Using the assumption of average uniformity of the $m$-lattice and the symmetry $m\to -m$ of the model [see Eq.~\eqref{Ft:eqn}] we can conclude that $C(m)=0$ and that $D(m)$ does not depend on $m$.
	
	We then conclude that
	\begin{equation}\label{pap:eqn}
		\frac{d}{dt} p_m(t) = \sigma(t) D\, \partial_m^2 p_m(t)
	\end{equation}
	plus higher derivatives term, whose contribution goes to $0$ as the time goes on.
	
	
	
%
This is a diffusion equation. If we initialize with a $p_m(0)$ symmetric for $m\to -m$, the dynamics will preserve this symmetry. This is our case because we initialize with $p_m(0)=\delta_{m\,0}$. Therefore, the quantity $m_t^2\equiv\sum_m m^2 p_m(t)$ is the variance of the $p_m(t)$ distribution. Notice that $e(t)$ is proportional to $m_t^2$ being $e(t)=\frac{\kbar^2}{2}m_t^2$. It is well known that the variance of a distribution obeying a diffusion equation like Eq.~\eqref{pap:eqn} obeys the equation
\begin{equation}\label{diffo2:eqn}
\frac{\ud }{\ud t} m_t^2=\frac{1}{2}\sigma(t)D\,.
\end{equation}
Notice that if $\sigma(t)={\rm const.}$, we find a diffusive behaviour of the energy, with $e(t)\propto m_t^2$ linearly increasing in time, in agreement with the results known in literature for a driving undergoing a noise with properties invariant under time translations~\cite{klappauf1998observation}. 

In the case of our dynamics, Eq.~\eqref{sissi:eqn} is valid and then we need to estimate $\overline{F^2}(t)=\text{Var}\left[F(t)\right]$. \mf{Applying Eq.~\eqref{var-F-estimate:eq}, then we then have}
\begin{equation}\label{deppe:eqn}
  \frac{\ud}{\ud t}m_t^2\propto\frac{D}{m_t}\,.
\end{equation}
Solving this simple differential equation we find
\begin{equation}
  e(t)=\frac{\kbar^2}{2}m_t^2\propto D^{1/3}t^{2/3}\,.
\end{equation}
This time dependence has been observed for long times and $\epsilon\in[0.1,1]$ (see Fig.~\ref{semievo1:fig} and Ref.~\cite{kicked-rotors}).

\begin{figure}
 \begin{center}
  \begin{tabular}{c}
    \includegraphics[width=8cm]{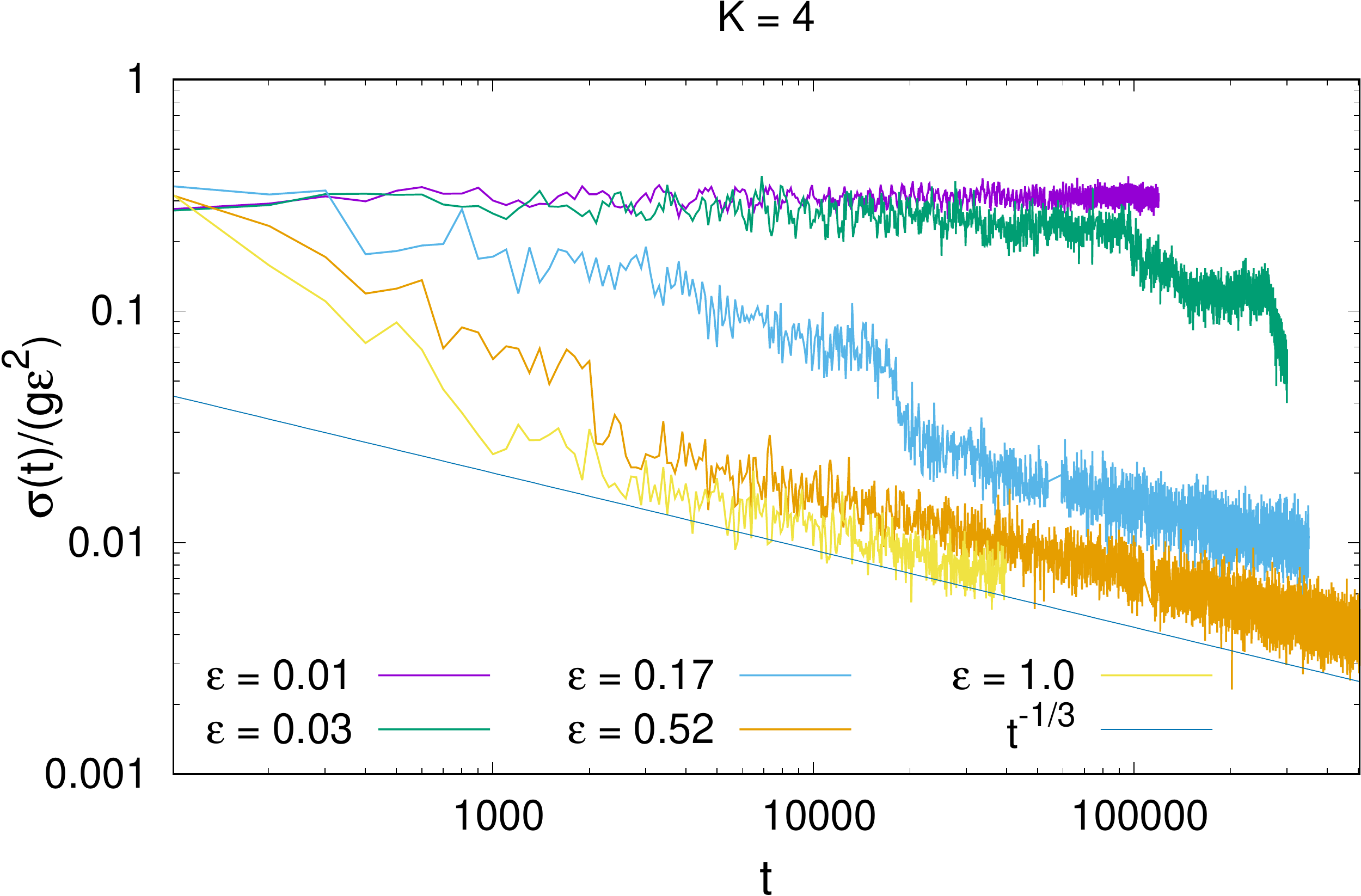}\\
    \includegraphics[width=8cm]{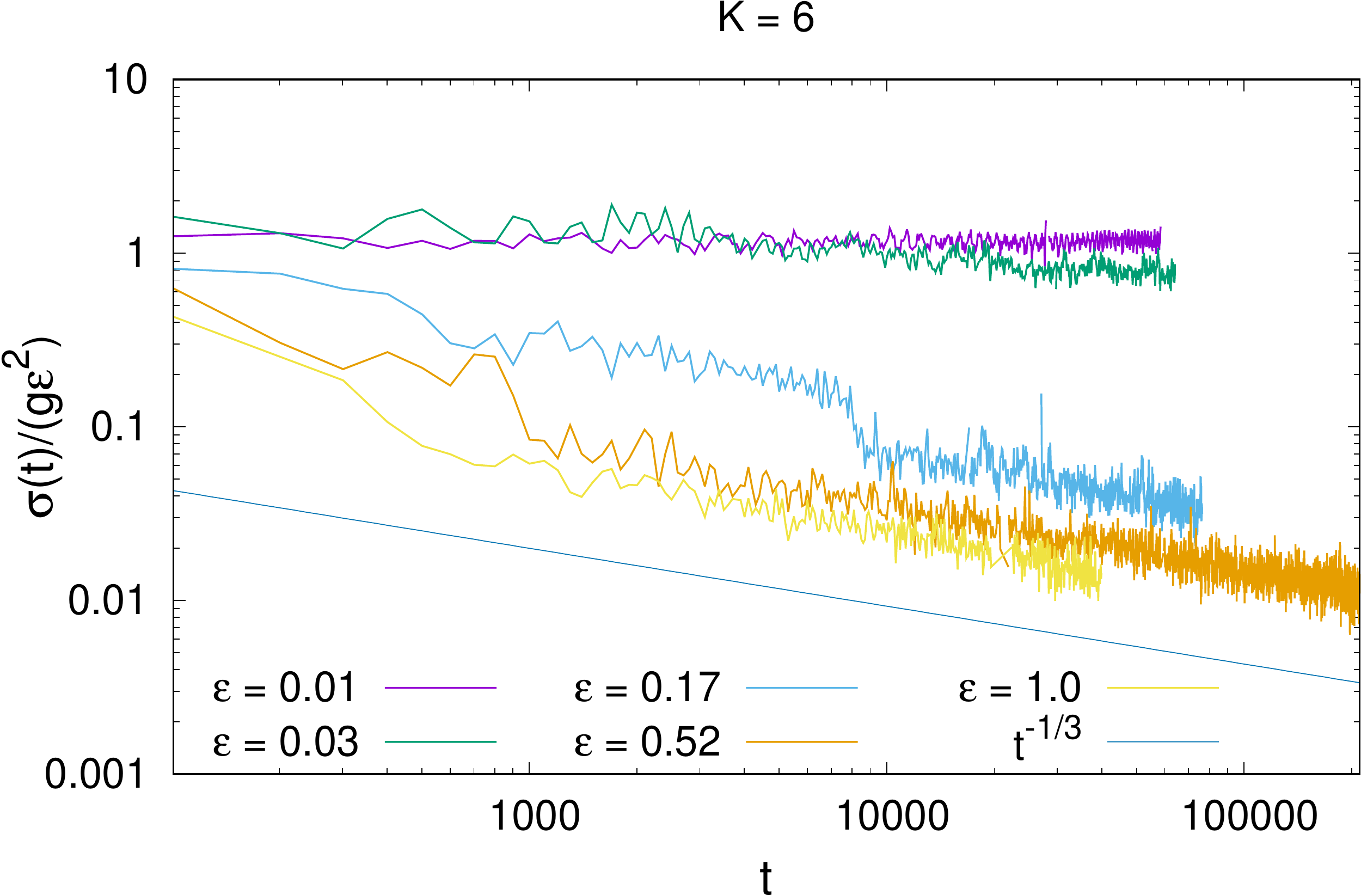}\\
    \includegraphics[width=8cm]{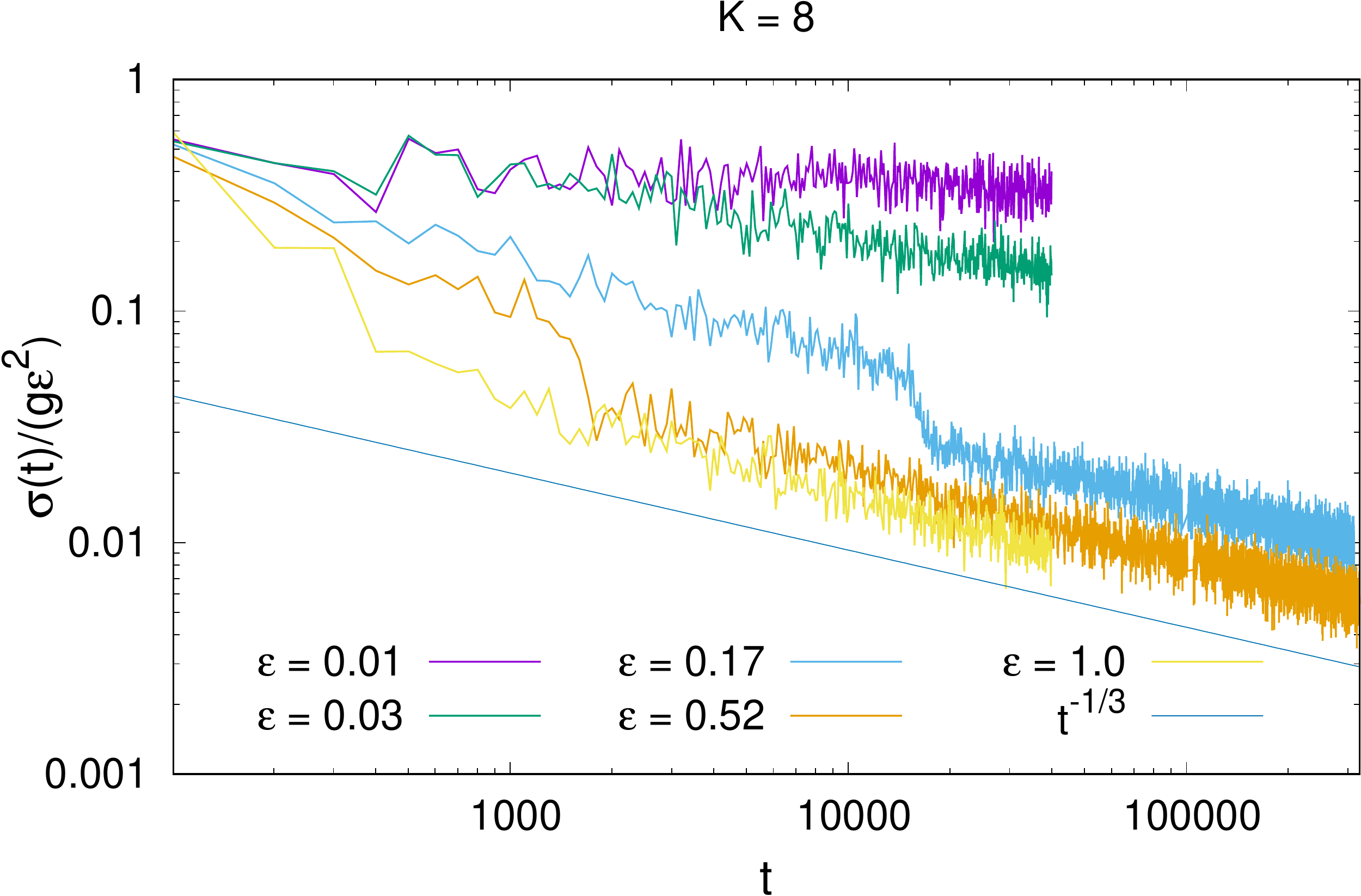}\\
  \end{tabular}
 \end{center}
 \caption{Bilogarithmic plot of $\sigma(t)/(g\epsilon^2)$ versus $t$ ($t$ is coarse grained with $\Delta t = 50$). Notice the power-law decrease consistent with the law $\sigma(t)\propto t^{-1/3}$ when $e(t)\sim t^{2/3}$ in Fig.~\ref{semievo1:fig}. Other numerical parameters: $\kbar = 1.7$, $M=20153$.}
    \label{F2emievo:fig}
\end{figure}

This theory gives rise to another prediction which we can numerically check. \mf{In fact, combining the scaling of the energy with Eq.~\eqref{var-F-estimate:eq}, we find}
\begin{equation}
  \sigma(t)\propto\overline{F^2}(t)\propto t^{-1/3}\,.
\end{equation}
Whenever there is power-law increase of the energy with exponent $2/3$, this is exactly the time dependence of $\sigma(t)$ that we numerically observe (see Fig.~\ref{F2emievo:fig}). So, $\sigma(t)$ decreases with time and then the coarse-graining approximation improves when larger and larger times are considered. This fact explains why the curves in Fig.~\ref{semievo1:fig} tend towards a power law with exponent $2/3$ at large times.

\mf{
We now discuss when we expect our analysis to break down. This is an important point, for instance fitting the curves for $\epsilon=0.52$ in Fig.~\ref{semievo1:fig} for $t\geq 3000$ we find a value of the power-law exponent $\gamma\simeq 0.72$. Starting from smaller values of $t$ one would get even larger values of $\gamma$. The point is that the curves slowly tend towards the power-law behaviour with exponent $2/3$. The reason is that, as already stressed, our analysis heavily relies on the assumption $\sigma(t)\ll1$. Before this condition is valid, in the case $\epsilon=0.52$, one must wait a very long transient time. Plugging Eq.~\eqref{var-F-estimate:eq} into the definition of $\sigma(t)$, we have that 
\begin{equation}
	\sigma(t) \sim \frac{\epsilon^2 K^4}{m_t}.
\end{equation}
Thus, for $\sigma(t)\ll1$ to hold we must have $m_t \gg {\epsilon^2 K^4}$. This implies that for large $K$, there could be a very long transient regime where the energy absorption could follow a different scaling in time. This could explain the linear energy increase found in~\cite{kicked-rotors} for large $K$ and $\epsilon$, which might be just a transient behaviour. For $\epsilon\ll 1$, in contrast, the chaotic behaviour of the system becomes apparent after a transient exponentially large in $1/\epsilon$, as a consequence of the Nekhoroshev theorem~\cite{Nekhoroshev1971} (see Sec.~\ref{nekhoroshev:sec}). In this transient the system looks like integrable and shows dynamical localization. This fact could explain the dynamical localization without energy increase that we observe in Figs.~\ref{semievo1:fig} and~\ref{F2emievo:fig} for $\epsilon=0.01$. In order to see if at some point an energy increase with exponent $2/3$ starts, one needs a time of the order of some power of $\exp(1/\epsilon^b)$ for some $b>0$. For instance, taking $b=1$ one gets $\exp(1/\epsilon)=\exp(10^2)\simeq 2.6\times 10^{43}$, which is beyond our numerical capabilities.
}

\section{Largest Lyapunov exponent} \label{lyapunov:sec}
Chaos was relevant in the last section in order to treat $F(t)$ as an effective noise.
We concentrate here in characterizing the chaotic properties of this model and we evaluate the largest Lyapunov exponent~\cite{Ott:book} $\lambda(\mathcal{T})$ as a stroboscopic average over $\mathcal{T}$ periods, and study its convergence with $\mathcal{T}$, using the method explained in~\cite{PhysRevA.14.2338}. We compare two trajectories with nearby initializations, one with $\beta_0(0)=1$, $\beta_{m\neq 0}(0)=0$, another with $\beta_0'(0)=\sqrt{1-\delta^2}$, $\beta_1(0)=\delta$, $\beta_{m\neq 0,\,1}=0$. We choose $\delta=10^{-10}$ and we study the rate of exponential divergence in each period in the following way. At the end of the first period the initial distance $\sqrt{2}\delta$ has become $d_1$. Before evolving over another period, we leave $(\ldots\,\beta_m(n)\,\ldots)$ unchanged and modify $(\ldots\,\beta_m'(n)\,\ldots)$: We move it along the line joining it with $(\ldots\,\beta_m(n)\,\ldots)$ so that the distance between the two trajectories become again $\delta$. We iterate this procedure over $\mathcal{T}$ periods obtaining the sequence of distances $d_1$, $d_2$, $\ldots$, $d_{\mathcal{T}}$. The largest Lyapunov exponent is evaluated as the average rate of exponential divergence 
\begin{equation}
  \lambda(\mathcal{T})=\frac{1}{\mathcal{T}}\sum_{j=1}^\mathcal{T}\log\left(\frac{d_j}{\sqrt{2}\delta}\right)\,.
\end{equation} 
This quantity  reaches a limit $\lambda$ over a finite $\mathcal{T}$ whenever the phase space is a bounded set.

\begin{figure}
 \begin{center}
  \begin{tabular}{c}
    \includegraphics[width=8cm]{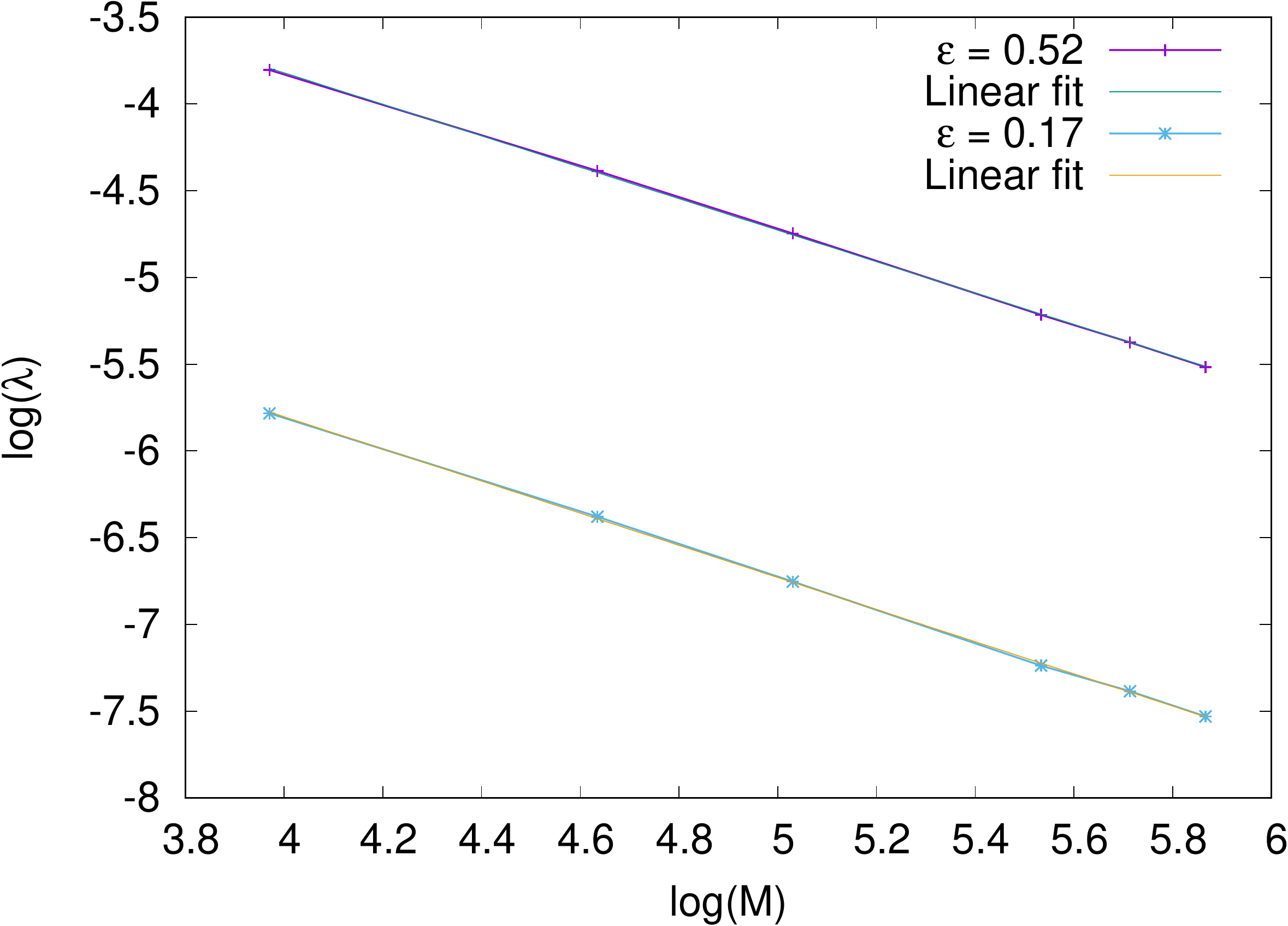}
  \end{tabular}
 \end{center}
 \caption{Largest Lyapunov exponent $\lambda$ versus $2M+1$. Numerical parameters: $K=6$, $\epsilon=0.52$, $\kbar = 1.7$, $\mathcal{T}\geq 10^6$.}
    \label{lyapunov:fig}
\end{figure}

The phase space is bounded for any finite $M$. For each choice of $M$ we take $\mathcal{T}$ so that we have reached convergence, and so we find that $\lambda$ decays as a power law with $2M+1$ (Fig.~\ref{lyapunov:fig}). We also fit $\log(\lambda)$ versus $\log(2M+1)$ with a straight line and the slope of the fitting line is $-0.91\pm 0.05$. Therefore, in the limit $M\to\infty$ the system should tend towards a regular behaviour with vanishing $\lambda$. 

\begin{figure}
 \begin{center}
  \begin{tabular}{c}
    \includegraphics[width=8cm]{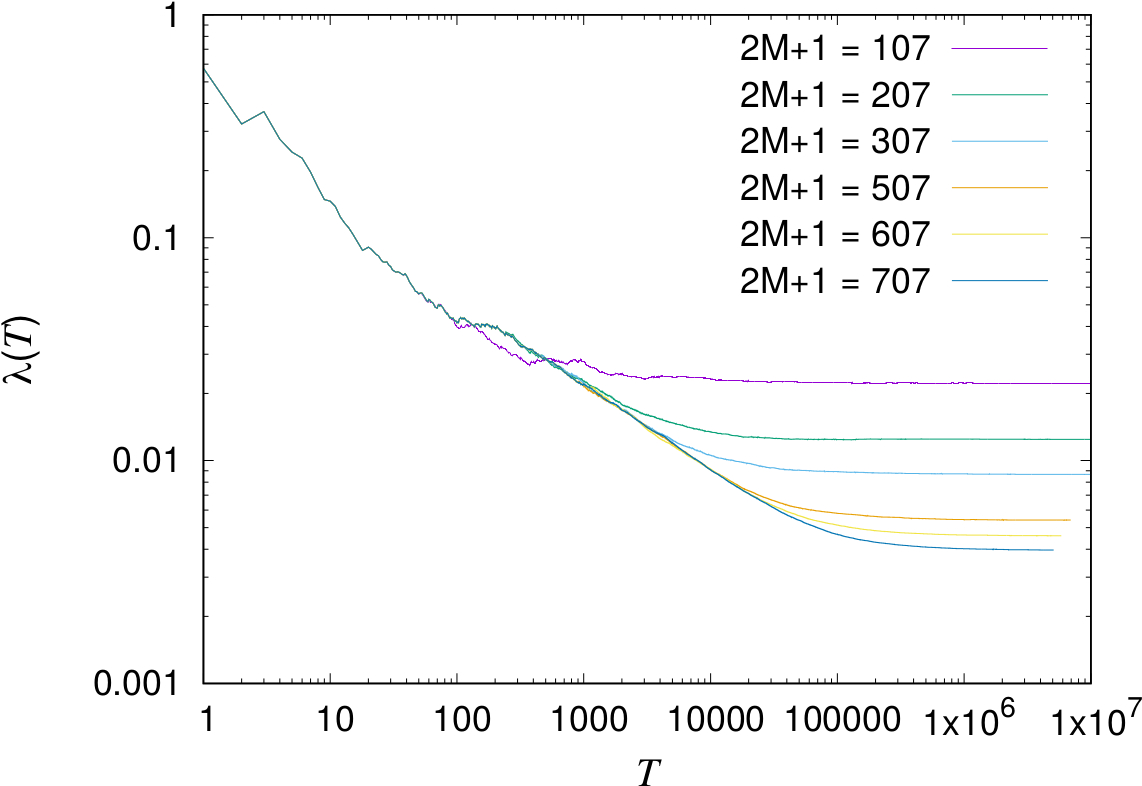}
  \end{tabular}
 \end{center}
 \caption{Largest Lyapunov exponent $\lambda(\mathcal{T})$ versus $\mathcal{T}$ for different values of $M$. Same numerical parameters as in Fig.~\ref{lyapunov:fig}}
    \label{lyapunov1:fig}
\end{figure}

This looks like a paradox for a thermalizing system (see Fig.~\ref{semievo:fig}), but we should not worry so much. 
We plot $\lambda(\mathcal{T})$ versus $\mathcal{T}$ for different values of the truncation $M$ in Fig.~\ref{lyapunov1:fig}. We see that $\lambda(\mathcal{T})$ decreases as a power law with $\mathcal{T}$, until it saturates to a plateau decreasing with $2M+1$ as a power law (see Fig.~\ref{lyapunov:fig}). 
For $M\to\infty$ the power law decay would last forever: the phase space is not bounded, the dynamics wanders away towards $m\to\infty$ and the system explores parts of the phase space with larger and larger $m$ and smaller and smaller exponential divergence of the trajectories. This looks reasonable: the ratio of the perturbation inducing chaos and the unkicked part of the Hamiltonian is $\epsilon/m^2$, so for large $m$ there is less chaos. 

For any finite $\mathcal{T}$, $\lambda(\mathcal{T})$ is nonvanishing because the dynamics is restricted to finite values of $m$ and gives a measure of the average Lyapunov exponent in that range of $m$. We can see that when you probe parts of the phase space with larger $m$, you get a smaller value of the Lyapunov exponent. We notice that $\lambda(\mathcal{T})$ relaxes to the plateau at the same time when the dynamics saturates to the maximum possible attainable value of $m$ and the energy in Fig.~\ref{semievo:fig} attains the $T=\infty$ value at finite $M$. 

\begin{figure}
 \begin{center}
  \begin{tabular}{c}
    \includegraphics[width=8cm]{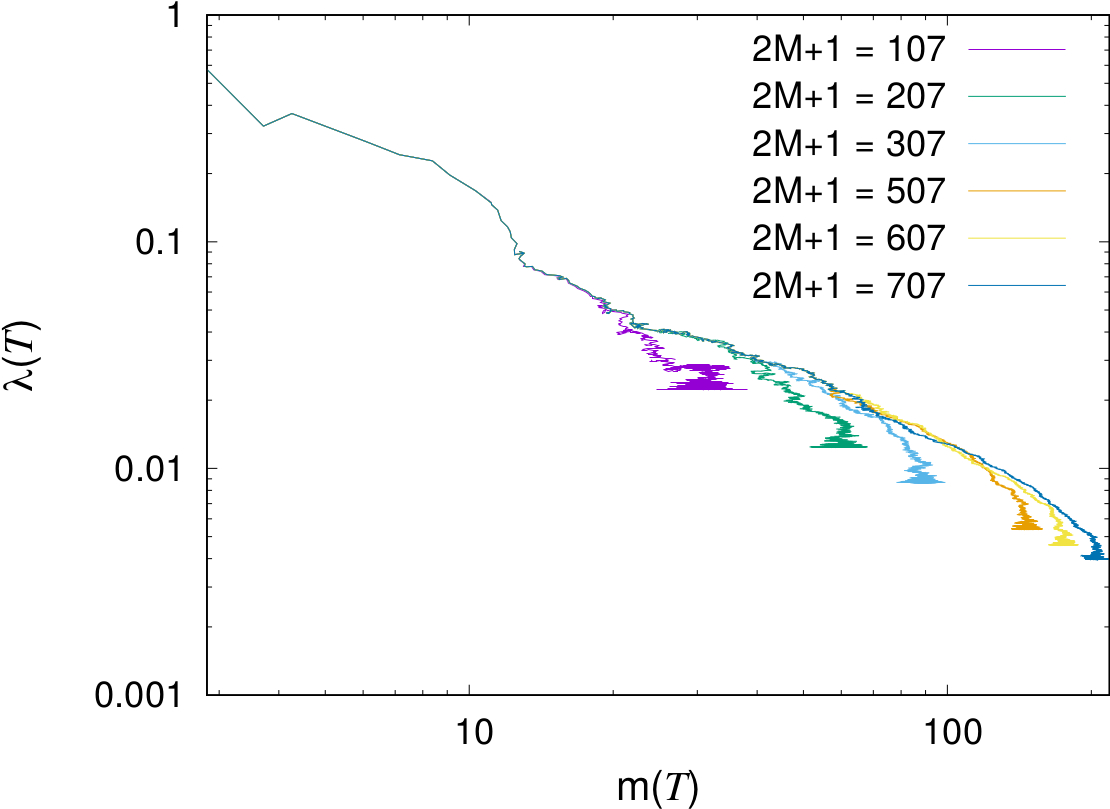}
  \end{tabular}
 \end{center}
 \caption{Largest Lyapunov exponent $\lambda(\mathcal{T})$ versus $m(\mathcal{T})\equiv\sqrt{2e(\mathcal{T})}/\kbar$ for different values of $M$. Same numerical parameters as in Fig.~\ref{lyapunov:fig}}
    \label{lyapunov2:fig}
\end{figure}

\begin{figure}
 \begin{center}
  \begin{tabular}{c}
    \includegraphics[width=8cm]{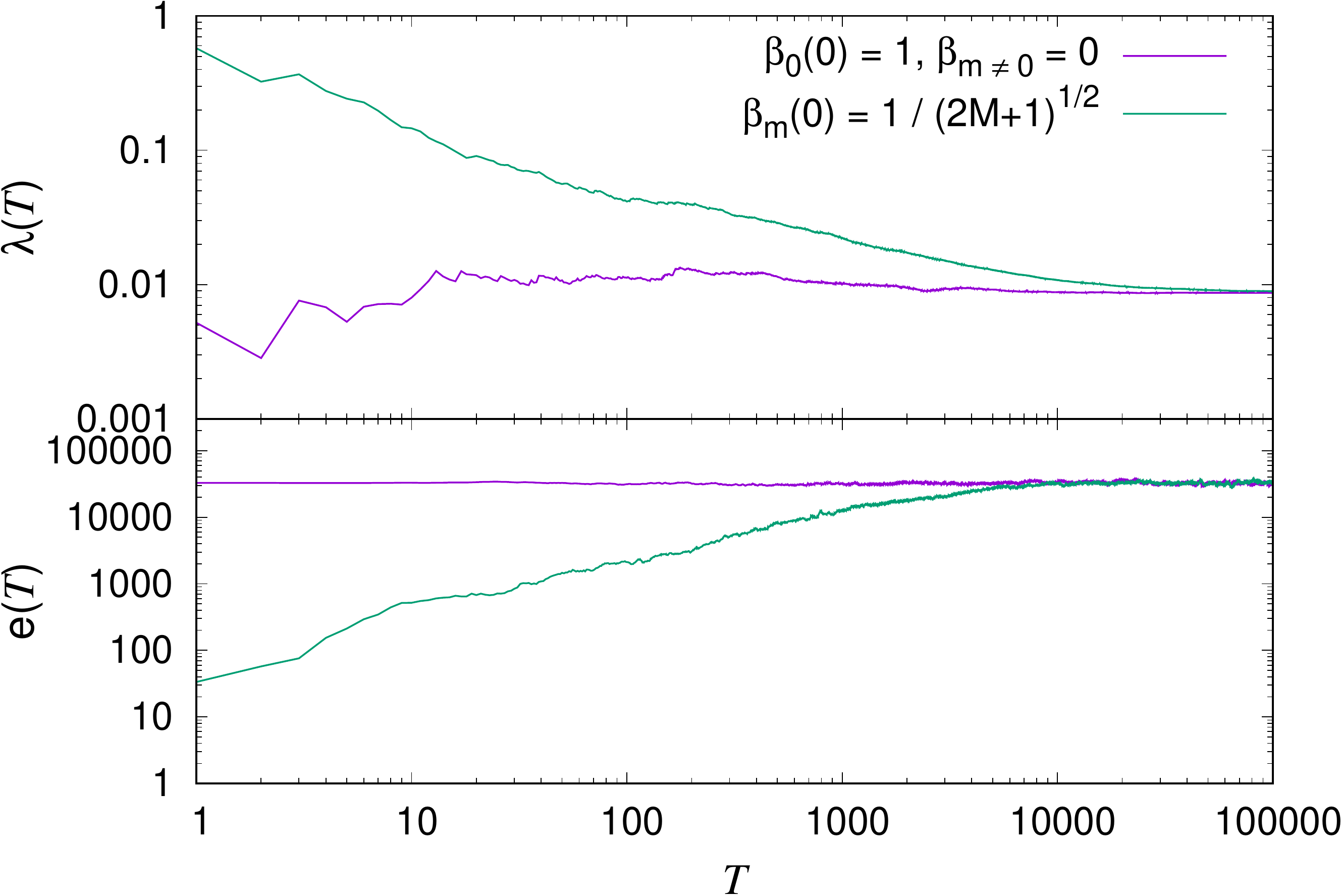}
  \end{tabular}
 \end{center}
 \caption{Largest Lyapunov exponent $\lambda(\mathcal{T})$ versus $\mathcal{T}$ for different initializations (upper panel) and corresponding energy evolution (lower panel). Numerical parameters: $ M = 153,\, K = 4,\, \epsilon= 0.52,\, \kbar = 1.7$.}
    \label{lyapunov3:fig}
\end{figure}

An important information is that different time regimes in Fig.~\ref{lyapunov1:fig} correspond to different ranges of $m$: The larger $\mathcal{T}$, the larger $m$, the smaller the largest Lyapunov exponent $\lambda(\mathcal{T})$. We can see this fact more explicitly if we plot $\lambda(\mathcal{T})$  versus $m(\mathcal{T})\equiv\sqrt{2e(\mathcal{T})}/\kbar$, as we do in Fig.~\ref{lyapunov2:fig}. We find further confirmation of this fact if we evaluate the largest Lyapunov exponent considering different initializations. Beyond the one considered up to now, we take another one where for one trajectory $\beta_m(0)=1/\sqrt{2M+1}$ $\forall\,m$ and for the other $\beta_{m\neq0\,1}'(0)=1/\sqrt{2M+1}$, $\beta_{m\neq0\,1}'(0)=\sqrt{1-\delta}/\sqrt{2M+1}$, $\beta_{m\neq0\,1}'(0)=\sqrt{1+\delta}/\sqrt{2M+1}$. In the second initialization scheme the energy expectation is equal to the $T=\infty$ value from the beginning and then the dynamics explores large values of $m$ from the beginning. We show an example of the situation in Fig.~\ref{lyapunov3:fig}. In the upper panel we show the evolution of $\lambda(\mathcal{T})$. We see that with the second initialization $\lambda(\mathcal{T})$ is small from the beginning, consistently with the fact that the relevant values of $m$ are larger and our assumption of a $\lambda_m$ decreasing with $m$. For both initializations $\lambda(\mathcal{T})$ reaches the same limit, and this occurs at the time where the energy has reached the $T=\infty$ value for both the initializations (lower panel of Fig.~\ref{lyapunov3:fig}).

\subsubsection{$N\gg 1$}
A positive Lyapunov exponent has an important implication for the dynamics at $N\gg 1$ but finite. The $\beta_m$ have fluctuations of order $1/\sqrt{N}$ has we can see in Eq.~\eqref{comm:eqn}. Because of chaotic dynamics, this initial uncertainty increases exponentially fast in time with a rate given by the Lyapunov exponent. The larger $m$, the smaller the Lyapunov exponent, with a dependence $\lambda_m$ resembling the one in Fig.~\ref{lyapunov2:fig}. So, the Gross-Pitaevskii equations Eq.~\eqref{evolama:eqn} are valid until the initial uncertainty becomes of order 1. This occurs for a time $t$ such that $\nep^{\lambda_m t}/\sqrt{N}\sim 1$, that's to say
\begin{equation}
  t\sim\frac{1}{2\lambda_m}\log N\,.
\end{equation}
For $m\to\infty$ our results suggest that $\lambda_m\to 0$, and so in that limit the Gross-Pitaevskii equations are true for a time tending to infinity, even for finite $N$.
\section{Conserved quantities and the Nekhoroshev theorem} \label{nekhoroshev:sec}
\subsection{Nekhoroshev theorem in a nutshell}
Nekhoroshev theorem~\cite{Nekhoroshev1971} applies when an integrable system is perturbed with a term breaking the integrability. A classical integrable system is such if it has as many integrals of motion $I_j$ as degrees of freedom and these integrals of motion are in involution, this gives to the dynamics peculiar regularity properties~\cite{Arnold:book}. Perturbing the integrable system with an integrability breaking term, these quantities are no more conserved and they depend on time: $I_j(t)$ deviate from their initial value. Let us call $\epsilon$ the strength of the perturbation. Nekhoroshev theorem says that there are two positive real numbers $a$, $b$ such that $|I_j(t)-I_j(0)|\leq \epsilon^{b}$ whenever
\begin{equation}\label{neko:eqn}
 t\leq \exp(C\epsilon^{-a})
\end{equation}
for some positive constant $C$. This is a very important information because it tells us that for $\epsilon\ll 1$ the $I_j(t)$ are approximately conserved for a time exponentially large in the inverse perturbation, that's why the Nekhoroshev theorem is also called ``classical prethermalization''. We are going to show that it is valid also for our Gross-Pitaevskii equations.
\subsection{Conserved quantities at $\epsilon=0$}\label{cocon:sec}
Let us start focusing on $\epsilon=0$. In this case, Eqs.~\eqref{evolama:eqn} coincide with the Schr\"odinger equation of the single kicked rotor in the momentum basis (see for instance~\cite{Haake:book}). Now we are going to show that this noninteracting model, if probed stroboscopically in time, shows infinitely many conserved quantities local in $m$. They can be constructed by means of the Floquet diagonalization of Eq.~\eqref{evolama:eqn} at $\epsilon=0$. This procedure is equivalent to evaluating the single-rotor Floquet modes~\cite{Shirley_PR65,Samba} in momentum basis. To find them, one gets from Eq.~\eqref{ultron:eqn} the eigenvalue equation $\hat{U}_0(1)\ket{\phi_j(0)}=\nep^{-i\mu_j t}\ket{\phi_j(0)}$ and solves it, expanding $\hat{U}_0(1)$ and $\ket{\phi_j(0)}$ in the momentum eigenbasis. More specifically, we are focusing here on the time immediately after the kick, so we consider the form $\hat{U}_0(1^+)=\nep^{-iK\cos\hat{\theta}}\nep^{-i\hat{p}^2/(2\kbar)}$. 
%
%
%
%
The eigenvalues are of the form $\nep^{-i\mu_j}$; the eigenvectors are the Floquet modes and they have the form ${\bf V}_j\equiv\left(\begin{array}{ccccc}\ldots&U_{m-1\,j}&U_{m\,j}&U_{m+1\,j}&\ldots\end{array}\right)^T$ (in the notation of Sec.~\ref{subdiffusion:sec} we have $\ket{\phi_j(0)}=\sum_m\,U_{m\,j}\ket{m}$). 

From one side we see that if we prepare the system in the condition $\beta_m(0)=U_{m\,j}$ $\forall\,m$, the evolution reduces to $\beta_m(n)=\nep^{-in\mu_j}U_{m\,j}$ being ${\bf V}_{j}$ a Floquet mode of the single-rotor dynamics. From the other side, we see that the quantity
\begin{equation}\label{olpha:eqn}
  O_j(n)=\sum_{m=-\infty}^\infty U_{m\,j}^*\,\beta_m(n)
\end{equation}
with $n$ integer evolves as $O_j(n)=\nep^{i\mu_j n}O_j(0)$, so that $|O_j|^2$ is conserved, whatever are the initial conditions in the $\beta_m$. We notice that the $|O_j|$ are local in $m$ due to the Anderson localization of the Floquet states~\cite{PhysRevLett.49.509,PhysRevA.29.1639,kicked-rotors} Moreover, we can show that the $|O_j|$ are in involution, that's to say their Poisson bracket vanishes. This can be easily seen by evaluating the Poisson bracket $\{|O_j|^2,\,|O_{j'}|^2\}$ and showing that it vanishes by using the elementary Poisson brackets Eq.~\eqref{brackets:eqn} and the orthonormality condition $\sum_m U_{m\,j}^* U_{m\,j'}=\delta_{j\,j'}$.

 This is an important remark because it implies that the $\epsilon=0$ system is an integrable system in the classical Hamiltonian sense~\cite{Arnold:book}. From one side the integrals of motion are as many as the degrees of freedom (imposing a truncation one sees that the $j$ are as many as the $m$). From the other side the Poisson brackets of the integrals of motion vanish so that they are in involution. So, if we apply a perturbation to it, the Nekhoroshev theorem should be valid, as we are going to explicitly show now.

%
\subsection{Nekhoroshev estimate for $\epsilon>0$}
\begin{figure}
 \begin{center}
  \begin{tabular}{c}
    \vspace{0.5cm}\\\includegraphics[width=8cm]{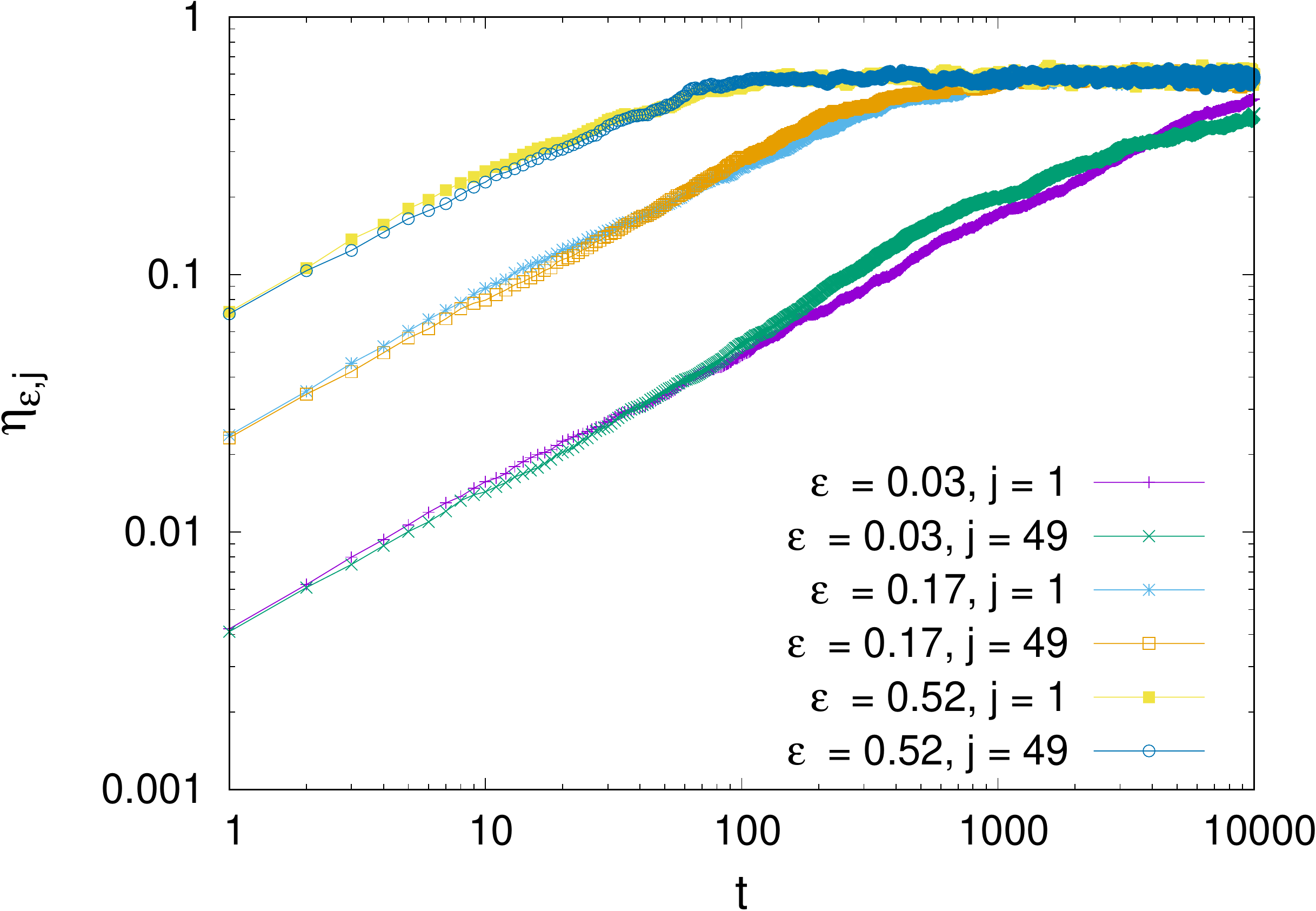}
  \end{tabular}
 \end{center}
 \caption{$\eta_{j,\,\epsilon}(t)$ versus $t$ for different values of $j$ and $\epsilon$.  Numerical parameters: $K=6$, $\kbar = 1.7$, $M=153$, $N_{\rm diso}=624$.}
    \label{distribution:fig}
\end{figure}

\begin{figure}
 \begin{center}
  \begin{tabular}{c}
    \vspace{0.5cm}\\\includegraphics[width=8cm]{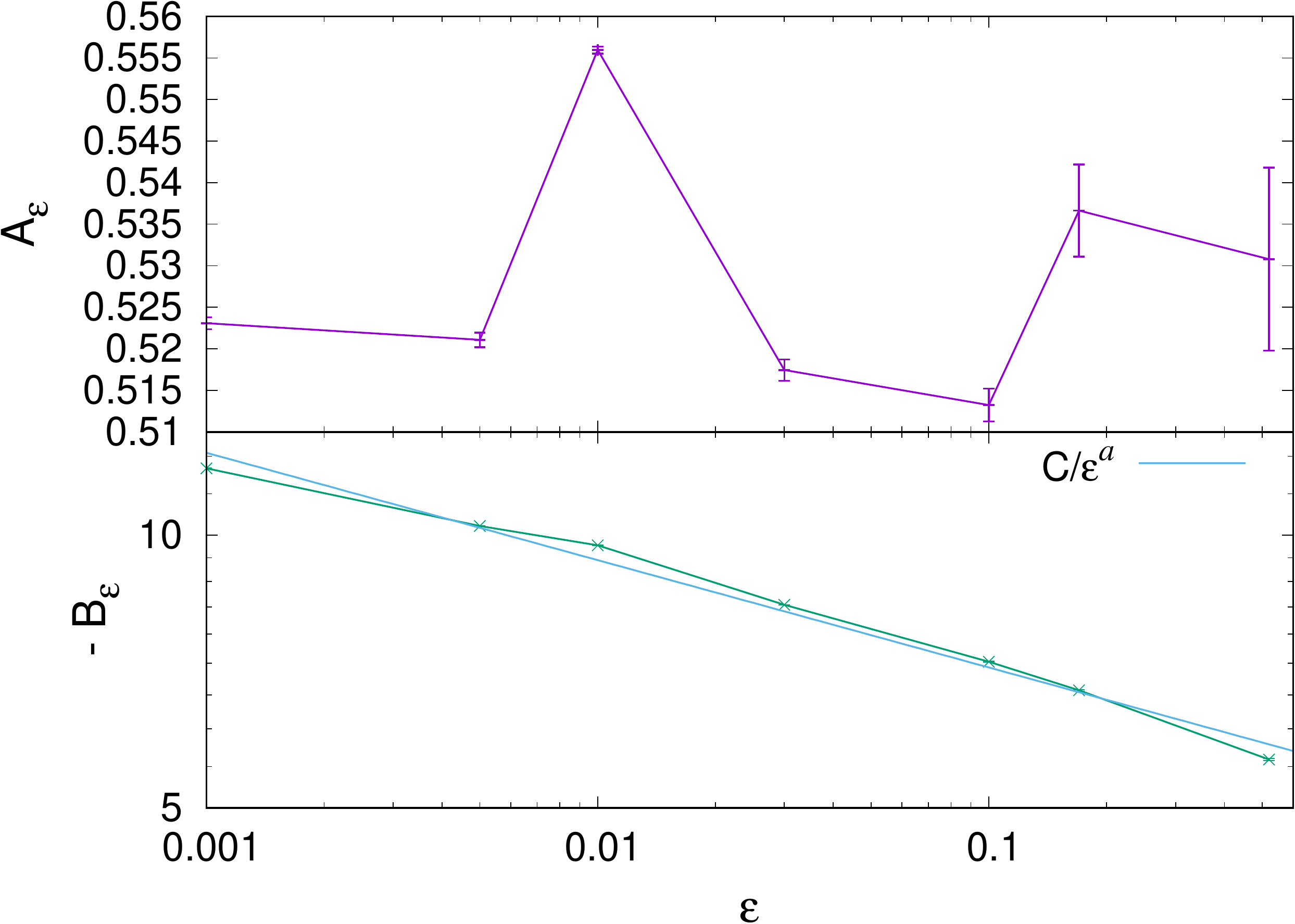}
  \end{tabular}
 \end{center}
 \caption{${A}_{\epsilon}$ and $-{B}_{\epsilon}$ versus $\epsilon$. The errorbars are evaluated as the averages over $j$ of the errorbars given by the fitting procedure: they are much larger than the standard deviations over $j$ of respectively $A_{j,\,\epsilon}$ and $B_{j,\,\epsilon}$. In the lower panel we plot also the result of the linear fit of the bilogarithmic plot. Numerical parameters: $K=6$, $\kbar = 1.7$, $M=153$, $N_{\rm diso}=624$.}
    \label{mediati_AB:fig}
\end{figure}

 It is very interesting to study the time dependence of $|O_j(t)|_{\epsilon>0}$ when $\epsilon>0$ is considered. Let us initialize the system with the condition $\beta_m(0)=1/\sqrt{2M+1}$. We consider
\begin{equation}
  \delta_{j,\,\epsilon}(t)=||O_j(t)|_{\epsilon>0}-|O_j|_{\epsilon=0}|\,.
\end{equation}
We study the evolution in time of this quantity. In order to get rid of any influence of the initial state, we average over $N_{\rm diso}$ initial conditions taken as $\beta_m(0)=\nep^{-i\phi_m}/\sqrt{2M+1}$ with $\phi_m\in[0,2\pi]$ uniformly distributed random variables and call the average $\mean{\delta_{j,\,\epsilon}}(t)$. We show some examples of evolution of
\begin{equation}
  \eta_{j,\,\epsilon}(t)\equiv\frac{\mean{\delta_{j,\,\epsilon}}(t)}{\mean{|O_j|_{\epsilon=0}}}
\end{equation}
in Fig.~\ref{distribution:fig}. We see a initial power-law increase followed by a saturation to a plateau. We can fit the power law by means of a linear fit of the bilogarithmic plot 
\begin{equation} \label{dipendenza:eqn}
  \log\eta_{j,\,\epsilon}(t)=A_{j,\,\epsilon}\log t+B_{j,\,\epsilon}\,.
\end{equation}
 We average both $A_{j,\,\epsilon}$ and $B_{j,\,\epsilon}$ over $j$ and get 
\begin{align}
  {A}_{\epsilon}&=\frac{1}{2M+1}\sum_{j=1}^{2M+1}A_{j,\,\epsilon}\nonumber\\
  {B}_{\epsilon}&=\frac{1}{2M+1}\sum_{j=1}^{2M+1}B_{j,\,\epsilon}\,.
\end{align}
We plot ${A}_{\epsilon}$ and ${B}_{\epsilon}$ versus $\epsilon$ in Fig.~\ref{mediati_AB:fig}. We see two important properties which will be important in the following. The first one is that ${A}_{\epsilon}$ is almost constant in $\epsilon$ and equals $\sim 0.53$. The second one, emphasised by the bilogarithmic plot, is that $-{B}_{\epsilon}$ decays in $\epsilon$ as a power law, 
\begin{equation}\label{decay:eqn}
  -{B}_{\epsilon}\sim C/\epsilon^{a}\quad \text{with}\quad a = -0.118 \pm 0.006\ldots\,.
\end{equation}
where $a$ has been numerically obtained through the linear fit of the bilogarithmic plot.
Being the errorbars quite small in relation to the values of ${A}_{\epsilon}$ and ${B}_{\epsilon}$, we can focus on an average value of $\eta$ defined as
\begin{equation}\label{dipendenza1:eqn}
  \log{\eta}_{\epsilon}(t)\equiv{A}_{\epsilon}\log t+{B}_{\epsilon}\,.
\end{equation}
We give an estimate of the time when the deviation from the initial value gets significant enough. Consistently with the Nekhoroshev analysis, we estimate this time $t^*$ as the time when the condition $\eta_\epsilon(t^*)=\epsilon^b$ is verified. If $b>0$ is large enough, so that the resulting $t^*$ is still in the regime of power-law increase, using Eq.~\eqref{dipendenza:eqn} we get the result
\begin{equation}
  t^*=\epsilon^{b/{A}_\epsilon}\nep^{-{B}_\epsilon/{A}_{\epsilon}}\,.
\end{equation}
We have noticed that ${A}_\epsilon\simeq 0.53$. So, let us choose any $b>0.6$. Moreover, taking a $C'$ slightly larger than $C$ in Eq.~\eqref{decay:eqn}, we have $-{B}_{\epsilon}\leq C'/\epsilon^{a}$, as we can easily see in the lower panel of Fig.~\ref{mediati_AB:fig}. In this way we have that $\eta_\epsilon(t)$, the average deviation of the $O_j$ from the $\epsilon=0$ conserved values, is smaller or equal than $\epsilon^b$ when
\begin{equation}\label{Nekhoroshev:eqn}
  t^*\leq \epsilon^{b/{A}_\epsilon}\, \nep^{C''/\epsilon^a} \leq\nep^{C''/\epsilon^a}
\end{equation}
where we have used Eq.~\eqref{decay:eqn} and defined $C''=C'/0.52$. Eq.~\eqref{Nekhoroshev:eqn} is exactly the relation predicted by the Nekhoroshev theorem Eq.~\eqref{neko:eqn}.
\section{Conclusions}\label{conc:sec}
In conclusion we have studied the dynamics of the infinite-range coupled quantum kicked rotors. By mapping it over a model of interacting bosons we have performed exact diagonalization for quite large system sizes and truncations. We have analyzed the average level spacing ratio and we have found a generalized tendency towards ergodicity for increasing system sizes $N$, in agreement with previous analytical demonstrations. 

Then we have moved to the thermodynamic limit where the model is described by a system of Gross-Pitaevskii equations which reduces to the Schr\"odinger equation of the non-interacting model when the magnitude $\epsilon$ of the interaction term vanishes. For $\epsilon\neq 0$, these equations are equivalent to the dynamics of a single-rotor nonlinear effective Hamiltonian. This system gives rise to a power-law increase of the energy in time with exponent $\gamma\sim 2/3$ in a \mf{wide} range of parameters. We have been able to analytically explain this exponent by using a master equation approach based on the noisy behaviour of the nonlinear modulation of the kicking in the effective Hamiltonian. We have also applied a coarse graining in time and in momentum space, and thanks to the localization of the single-rotor Floquet states we were able to write a diffusion equation for the occupation probabilities of the momentum eigenstates whose solution gave rise to the exponent $2/3$. Moreover, we have predicted that the nonlinear modulation of the kicking, squared and coarse-grained in time, showed a power-law time dependence of the form $\sim t^{-1/3}$ which we have numerically verified. Remarkably, considering a the kick modulated by a noise with properties invariant under time translation, we could get the diffusion of the energy found in~\cite{klappauf1998observation}.

 We have shown that the Gross-Pitaveskij equations we have found are equivalent to a previously existing mean-field analysis of this model and we have studied chaos in them by means of the largest Lyapunov exponent. This quantity is a measure of chaos being an estimate of the rate of exponential divergence of nearby trajectories. We find that it decreases towards zero as a power law when looking to portions of the phase space with larger and larger momentum. From this fact follows that, for finite $N$, the time of validity of the Gross-Pitaevskii equations diverges with the momentum. Indeed, we have shown that this time is proportional to $\log N$ and inversely proportional to the rate of exponential divergence of the trajectories.

Later, we have considered the limit of $\epsilon=0$ and used Floquet theorem to construct integrals of motion of the stroboscopic dynamics. These integrals of motion are as many as the degrees of freedom and are in involution with each other, so the system is integrable in a classical sense. Taking $0<\epsilon\ll 1$ this integrability is broken and the integrals of motion deviate from their initial value. We have verified that they do so in a timescale exponentially large in the inverse $\epsilon$ and consistent with the predictions of the Nekhoroshev theorem. 

\mf{Future perspectives include the study of the model where the infinite-range interactions are replaced by long-range power-law interactions.
In this case the full permutation symmetry is broken and a strong qualitative change of the Hilbert space structure occurs~\cite{russomanno2020longrange} which might turn the subdiffusion into diffusion. On the other hand, one might expect that the $2/3$-exponent power-law behaviour is preserved, as long as the mean-field description is valid and chaos still gives rise to local correlations in time. This long-range case could be numerically investigated by means of a matrix product
state description and a time-dependent variational
principle (MPS-TDVP, used e.g. in~\cite{Silva} for a different power-law interacting model)}.
Moreover, it will be interesting to consider the case of nonuniform initialization, and see if the dynamics is described by a system of equations of the nonlinear Schr\"odinger equation form Eq.~\ref{evolama2:eqn}, which might still lead to energy subdiffusion. Other prospects of future work will be the analysis 
of other invariant subspaces of the Hamiltonian, the application of this analysis to the infinite-range coupled version of the kicked Bose-Hubbard chain~\cite{Michele_arxiv}, and the extension of the master-equation approach to other models.
\acknowledgements{A.~R. thanks D.~Rossini and A.~Tomadin for the access to the late GOLDRAKE cluster, where part of the numerical analysis for this project was performed.}
\appendix
\section{Bosonic representation}\label{boson:sec}
In this appendix we discuss the bosonic mapping used in the text. This mapping was invented by~\cite{federica} for a similar infinite-range coupled model. Let us consider a system of $N$ rotors, and we explicitly work out the representation of the operator
\begin{equation}
 \hat{\mathcal{O}}=\sum_{l=1}^N\nep^{-i\hat{\theta}_l}
\end{equation}
(for the other operators the procedure is very similar). We consider the basis $\{\ket{m}_l\}$ of eigenstates of the operator $\hat{p}_l$ with eigenvalues $\kbar m$ with $m\in\mathbb{Z}$. In this basis, the operator $\hat{O}$ reads
\begin{equation}\label{opi:eqn}
  \hat{\mathcal{O}}=\sum_{l=1}^N\sum_{m=-\infty}^\infty\ket{m+1}_l\hspace{-1cm}{\phantom{\ket{m+1}}_l}\bra{m}\,.
\end{equation}
Because the system is fully symmetric under permutations, we can restrict to the states even under all the possible $N!$ permutations. If we call $\hat{P}$ the sum of all the permutation operators, we can take as basis of our Hilbert space the states
\begin{widetext}
\begin{equation} \label{defiopi:eqn}
  \ket{\ldots\, n_{m-1}\,n_m\,n_{m+1}\,\ldots}\equiv\frac{1}{\sqrt{N!\,\prod_{m=-\infty}^\infty n_{m}!}}\hat{P}\ket{\ldots\,\underbrace{(m-1\ldots m-1)}_{n_{m-1}}\,\underbrace{(m\ldots m)}_{n_{m}}\underbrace{(m+1\ldots m+1)}_{n_{m+1}}\,\ldots}\,.
\end{equation}
\end{widetext}
There are $N$ rotors, so $\sum_{m=-\infty}^\infty n_m=N$. The factor in front is for normalization, you divide by $\sqrt{N!}$ because there are $N!$ possible permutations. And you divide also by all the $\sqrt{n_m!}$. Consider for instance $m=1$. Fixing everything else, there are $n_1!$ ways of rearrange the sites with $m=1$ and this increases the norm by a factor $n_1!$. You divide by $\sqrt{n_1!}$ and the norm is again 1.

Consider for instance the application of the operator $\hat{O}$ [Eq.~\eqref{opi:eqn}]. You find
\begin{widetext}
\begin{equation} \label{defio:eqn}
  \hspace{-1cm}\hat{O}\ket{\ldots\, n_{m-1}\,n_m\,n_{m+1}\,\ldots}=\frac{1}{\sqrt{N!\,\prod_{m=-\infty}^\infty n_{m}!}}\hat{P}\sum_{l=1}^N\sum_{m'=-\infty}^\infty\ket{m'+1}_l\hspace{-1.1cm}{\phantom{\ket{m+1}}_l}\bra{m'}\ket{\ldots\,\underbrace{(m-1\ldots m-1)}_{n_{m-1}}\,\underbrace{(m\ldots m)}_{n_{m}}\underbrace{(m+1\ldots m+1)}_{n_{m+1}}\,\ldots}\,.
\end{equation}
Using permutation symmetry, it is quite immediate to see that
\begin{equation} \label{defio:eqn}
  \hat{\mathcal{O}}\ket{\ldots\, n_{m-1}\,n_m\,n_{m+1}\,\ldots}=\frac{1}{\sqrt{N!\,\prod_{m=-\infty}^\infty n_{m}!}}\hat{P}\sum_{l=1}^N\sum_{m'=-\infty}^\infty n_{m'}\ket{\ldots\,\underbrace{(m'\ldots m')}_{n_{m'}-1}\,\underbrace{(m'+1\ldots m'+1)}_{n_{m'+1}}\underbrace{(m'+2\ldots m'+2)}_{n_{m'+2}}\,\ldots}\,.
\end{equation}
Slightly rearranging we find
\begin{equation} \label{defio:eqn}
  \hat{\mathcal{O}}\ket{\ldots\, n_{m-1}\,n_m\,n_{m+1}\,\ldots}=\hat{P}\sum_{l=1}^N\sum_{m'=-\infty}^\infty \frac{\sqrt{n_{m'}(n_{m'+1}+1)}\ket{\ldots\,\underbrace{(m'\ldots m')}_{n_{m'}-1}\,\underbrace{(m'+1\ldots m'+1)}_{n_{m'+1}}\underbrace{(m'+2\ldots m'+2)}_{n_{m'+2}}\,\ldots}}{\sqrt{N!\,\prod_{m=-\infty}^{m'-1}n_{m}! (n_m'-1)!(n_{m'+1}+1)!\prod_{m=m'+2}^\infty n_{m}!}}\,.
\end{equation}
From a formal point of view this is entirely equivalent to write
\begin{equation}
\hat{\mathcal{O}}\ket{\ldots\, n_{m-1}\,n_m\,n_{m+1}\,\ldots}=\sum_{m'=-\infty}^\infty\opbdag{m'+1}\opb{m'}\ket{\ldots\, n_{m-1}\,n_m\,n_{m+1}\,\ldots}
\end{equation}
where $\opb{m'}$ are bosonic operators and $\ket{\ldots\, n_{m-1}\,n_m\,n_{m+1}\,\ldots}$ is an eigenstate of the corresponding number operators $\hat{n}_{m'}$ with eigenvalues $n_{m'}$. In a fully analogous way one can show that
\begin{align}
  &\sum_{l=1}^N\nep^{i\hat{\theta}_l}\ket{\ldots\, n_{m-1}\,n_m\,n_{m+1}\,\ldots}=\sum_{m'=-\infty}^\infty\opbdag{m'}\opb{m'+1}\ket{\ldots\, n_{m-1}\,n_m\,n_{m+1}\,\ldots}\nonumber\\
  &\sum_{l=1}^N\hat{p}_l^2\ket{\ldots\, n_{m-1}\,n_m\,n_{m+1}\,\ldots}=\kbar^2\sum_{m'=-\infty}^\infty{m'}^2\hat{n}_{m'}\ket{\ldots\, n_{m-1}\,n_m\,n_{m+1}\,\ldots}\,.
\end{align}
In the subspace even under all the permutation symmetries, we can indeed write the two parts of the Hamiltonian Eq.~\eqref{H_a:eqn} as
\begin{align}\label{mappings:eqn}
&\frac{1}{2}\sum_{l=1}^N \hat{p}_l^2=\frac{\kbar^2}{2}\sum_{m'=-\infty}^\infty{m'}^2\hat{n}_{m'}\nonumber\\
&V(\hat{\boldsymbol\theta})=\frac{\kbar K}{2}\Big[\sum_{m=-\infty}^\infty \left(\opbdag{m}\opb{m+1}+{\rm H.~c.}\right)-\frac{\epsilon}{2(N-1)}\hspace{-0.1cm}\sum_{m,m'=-\infty}^\infty\hspace{-0.4cm}(\opbdag{m+1}\opb{m}\opbdag{m'}\opb{m'+1}+{\rm H.~c.})\Big]\,,
\end{align}
\end{widetext}
where $V(\hat{\boldsymbol\theta})$ is defined in Eq.~\eqref{H_sr:eqn}.
\section{From Eq.~\eqref{Om:eqn} to Eq.~\eqref{Om1:eqn}}\label{app:der}
We start considering the nested commutator in Eq.~\eqref{Om:eqn}. Expanding $\hat{V}$ in the Floquet-mode basis as $\hat{V}=\sum_{k\,l}V_{k\,l}(t)\ket{\phi_k(t)}\bra{\phi_l(t)}$ with $V_{k\,l}(t)$ time-periodic with period 1 we get
\begin{widetext}
\begin{align}
  \big[[\ket{\phi_i(0)}\bra{\phi_i(0)},\hat{V}_I(t)\big],\,\hat{V}_I(t)]&=\sum_{k\,l\,k'\,l'}V_{k\,l}(t)V_{k'\,l'}(t)\big[[\ket{\phi_i(0)}\bra{\phi_i(0)},\hat{U}_0^\dagger(t)\ket{\phi_k(t)}\bra{\phi_l(t)}\hat{U}_0(t)],\,\hat{U}_0^\dagger(t)\ket{\phi_{k'}(t)}\bra{\phi_{l'}(t)}\hat{U}_0(t)\big]\nonumber\\ &=
\sum_{k\,l\,k'\,l'}V_{k\,l}(t)V_{k'\,l'}(t)\hat{U}_0^\dagger(t)\big[[\ket{\phi_i(t)}\bra{\phi_i(t)},\ket{\phi_k(t)}\bra{\phi_l(t)}],\,\ket{\phi_{k'}(t)}\bra{\phi_{l'}(t)}\big]\hat{U}_0(t)\,,
\end{align}
where we have applied Eq.~\eqref{ultron:eqn}. Exploiting the orthonormality of the Floquet-mode basis we can evaluate in a lengthy but straightforward way the double commutator and get
\begin{align}
  \big[[\ket{\phi_i(0)}\bra{\phi_i(0)},\hat{V}_I(t)],\,\hat{V}_I(t)\big]&=
\sum_{k\,l}\hat{U}_0^\dagger(t)\Big\{V_{i\,k}(t)V_{k\,l}(t)\ket{\phi_i(t)}\bra{\phi_l(t)}-V_{k\,i}(t)V_{i\,l}(t)\ket{\phi_k(t)}\bra{\phi_l(t)}\nonumber\\
     &+V_{l\,i}(t)V_{k\,l}(t)\ket{\phi_k(t)}\bra{\phi_i(t)}-V_{k\,i}(t)V_{i\,l}(t)\ket{\phi_k(t)}\bra{\phi_l(t)}\Big\}\hat{U}_0(t)\nonumber\\
    &=
\sum_{k\,l}\Big\{V_{i\,k}(t)V_{k\,l}(t)\nep^{i(\mu_i-\mu_k)t}\ket{\phi_i(0)}\bra{\phi_l(0)}-V_{k\,i}(t)V_{i\,l}(t)\nep^{i(\mu_k-\mu_l)t}\ket{\phi_k(0)}\bra{\phi_l(0)}\nonumber\\
     &+V_{l\,i}(t)V_{k\,l}(t)\nep^{i(\mu_k-\mu_i)t}\ket{\phi_k(0)}\bra{\phi_i(0)}-V_{k\,i}(t)V_{i\,l}(t)\nep^{i(\mu_k-\mu_l)t}\ket{\phi_k(0)}\bra{\phi_l(0)}\Big\}\,.
\end{align}
\end{widetext}
At this point we perform the coarse-graining average. We average over $\Delta t$ which lasts many periods and we assume that the $\mu_k$ are incommensurate with the driving frequency $2\pi$~\cite{Russomanno_PRB11}. In this way we obtain
{\small
\begin{equation}
   \overline{\big[[\ket{\phi_i(0)}\bra{\phi_i(0)},\hat{V}_I(t)],\,\hat{V}_I(t)\big]}=2\sum_{k\neq i}\overline{|V_{k\,i}|^2}\ket{\phi_k(0)}\bra{\phi_k(0)}\,.
\end{equation}}
Here $\overline{|V_{k\,i}|^2}$ does not depend on the coarse-grained time because $|V_{k\,i}(t)|^2$ is periodic of period $1$ and the averaging time $\Delta t$ spans many periods. Substituting this expression into Eq.~\eqref{Om:eqn} we get Eq.~\eqref{Om1:eqn}.
%

\begin{thebibliography}{59}
\bibitem{noteb}
For a single rotor, or for $\epsilon=0$, the effective model Eq.~\eqref{hamo:eqn} is quadratic in the bosonic creation and destruction operators. Writing the Heisenberg equations for the destruction operators of the momentum occupations $i\kbar\frac{\ud}{\ud t}\hat{b}_{m\,H}(t)=[\hat{b}_{m\,H}(t),\,\hat{H}_b(t)]$ and writing the expansion $\hat{b}_{m\,H}(t)=\sum_{m''}\beta_{m\,m''}(t)\,\hat{b}_{m''}$ one obtains that the $\beta_{m\,m''}(t)$ obey exactly the Schr\"odinger equation for the single-kicked-rotor wave function in the momentum basis, which coincides with Eq.~\eqref{evolama:eqn} for $\epsilon=0$, identifying $\beta_{m\,m''}(t)$ with $\beta_m(t)$ in that equation. One can easily check that $\beta_{m\,m''}(t)=\braket{m|\hat{U}_0(t)|m''}$ [see the definition of $\hat{U}_0$ above Eq.~\eqref{interaction:eqn} and the definition of $\ket{m}$ in Sec.~\ref{subdiffusion:sec}] and equals the wavefunction in the momentum representation if the system is initialized in a momentum eigenstate.
\bibitem{notrasfo}
  Usually (see for instance~\cite{Sciolla_2}) one writes $\hat{Q}=\kbar x$ and $\hat{P}=-i\frac{\ud}{\ud x}$. The two descriptions are exactly equivalent due to the canonical transformation $\hat{P}=-\hat{\theta}$, $\hat{Q}=\hat{p}$.
\end{thebibliography}
\providecommand{\noopsort}[1]{}\providecommand{\singleletter}[1]{#1}%
\end{document}